\documentclass[12pt,a4paper]{article}
\usepackage{graphicx}
\usepackage{cite}

\makeatletter
\renewcommand{\@biblabel}[1]{}
\makeatother

\begin{document}

\newcommand{\EQ}{Eq.~}
\newcommand{\EQS}{Eqs.~}
\newcommand{\FIG}{Fig.~}
\newcommand{\FIGS}{Figs.~}
\newcommand{\TAB}{Tab.~}
\newcommand{\TABS}{Tabs.~}
\newcommand{\SEC}{Sec.~}
\newcommand{\SECS}{Secs.~}

\setlength{\baselineskip}{0.77cm}

\title{Controlling nosocomial infection 
based on structure of hospital social networks}
\bigskip

\author{Taro Ueno${}^{1}$ and Naoki Masuda${}^{2*}$\\
\ \\
\ \\
${}^1$ Tokyo Metropolitan Hiroo General Hospital,\\
2-34-10 Ebisu, Shibuya, Tokyo 150-0013, Japan\\
\ \\
${}^{2}$ 
Graduate School of Information Science and Technology,\\
The University of Tokyo,\\
7-3-1 Hongo, Bunkyo, Tokyo 113-8656, Japan
\ \\
$^*$ Author for correspondence (masuda@mist.i.u-tokyo.ac.jp)}

\maketitle

\newpage

\begin{abstract}
\setlength{\baselineskip}{0.77cm} Nosocomial infection (i.e. infection
in healthcare facilities) raises a serious public health problem, as
implied by the existence of pathogens characteristic to healthcare
facilities such as methicillin-resistant {\it Staphylococcus aureus}
and hospital-mediated outbreaks of influenza and severe acute
respiratory syndrome.  For general communities, epidemic modeling
based on social networks is being recognized as a useful tool.
However, disease propagation may occur in a healthcare facility in a
manner different from that in a urban community setting due to
different network architecture.  We simulate
stochastic susceptible-infected-recovered dynamics on social networks,
which are based on observations in a hospital in Tokyo, to explore
effective containment strategies against nosocomial infection.  The
observed social networks in the hospital have hierarchical and modular
structure in which dense substructure such as departments, wards, and
rooms, are globally but only loosely connected, and do not reveal
extremely right-skewed distributions of the number of contacts per
individual.  We show that healthcare workers, particularly medical
doctors, are main vectors (i.e. transmitters) of diseases on these
networks. Intervention methods that restrict interaction between
medical doctors and their visits to different wards shrink the final
epidemic size more than intervention methods that
directly protect patients, such as isolating patients in single
rooms. By the same token, vaccinating doctors with priority rather
than patients or nurses is more effective. Finally, vaccinating
individuals with large betweenness centrality (frequency of mediating
connection between pairs of individuals along the shortest paths) is
superior to vaccinating ones with large connectedness to others or
randomly chosen individuals, which was suggested by previous model studies.

\end{abstract}

\newpage

\section{Introduction}\label{sec:introduction}

Nosocomial infection, that is, infection in healthcare facilities, is
health-threatening for hospitalized patients and caregivers
\cite{Gastmeier05,Grundmann06}.  Pathogens such as
methicillin-resistant {\it Staphylococcus aureus} (MRSA) and
vancomycin-resistant enterococci (VRE) are mainly nosocomial in the
sense that the majority of cases are hospital-related
\cite{Salgado03,Cooper04bmj}.  In addition, as observed in recent
cases of influenza \cite{CDR05} and severe acute respiratory syndrome
(SARS) \cite{Ho03,Leung04,Svoboda04}, nosocomial infection serves as an
initiator or a booster of epidemic outbreaks at urban community and
world-wide levels. In accordance, mathematical models specialized in
nosocomial infections and their controls have been developed, with
intensive emphases on antibiotic-resistant bacteria
\cite{Lipsitch00,Pelupessy02,Forrester05,Bootsma06,Boldin07} and
intensive care units
\cite{Pelupessy02,Perencevich04,Forrester05,Boldin07}.  Quantitative
statistical models for nosocomial infection have also been reported
\cite{Cooper04bios,Forrester05,Forrester07}.

In this work we examine the possibility of using the information on
social networks in healthcare facilities for developing containment
protocols.  There have been many modeling reports on disease spreading
on complex social networks
\cite{Meyers03,Christley05,Keeling05,Meyers05,Pourbohloul05,Watts05,Colizza06,Green06,Riley07,Volz07}. A
central theoretical finding is that hubs, namely, those who have
contacts to relatively many others, communicate and enhance disease
transmission. Diseases spread to a great extent with larger probability
on heterogeneous networks such as scale-free networks in which the number
of contacts that each individual maintains is highly heterogeneous
\cite{Hethcote84,Anderson86,Albert00nat,Pastor01prl}.  Such
heterogeneous networks are operative in sexually transmitted diseases
\cite{Liljeros01} and computer viruses \cite{Pastor01prl}.

However, this and other findings in network epidemiology so far do not
seem to have a sufficient descriptive power for nosocomial infections
for two main reasons.  First, social networks of healthcare
facilities may be structured differently from networks of urban communities
or the whole world. In terms of size, a hospital is typically much
smaller than an urban community. In terms of microscopic structure, contact
patterns in a hospital are regulated by hierarchy defined by
departments, wards, and rooms.  To our knowledge, efficient
containment strategies for social networks with such modular structure
have not been established.  Second, a healthcare facility is composed
of individuals of distinct roles, such as patients, visitors, and
healthcare workers. Healthcare workers can be subdivided into
different classes, such as nurses and medical doctors.
Susceptibility, mortality, infectiousness, and many other factors that
affect how diseases spread depend on the type of individuals.  For
example, junior doctors may visit more wards than nurses do, possibly
carrying pathogens from ward to ward \cite{CDR05}.  Patients may be
less active but likely have larger case fatality (i.e. fraction of
death among infected individuals) than healthcare
workers \cite{Leung04,Forrester05}. In urban community social networks, the
role of different types of individuals in disease propagation may not
be so clear-cut.

In spite of seminal modeling work of nosocomial infection based on
network analysis \cite{Meyers03,Liljeros07}, 
how diseases spread in potentially hierarchical
networks of healthcare facilities composed of individuals of
different classes is not sufficiently understood.  In this work, we
model nosocomial infection in social networks based on observations of
contact frequency in a hospital in Tokyo.  We simulate stochastic SIR
epidemic dynamics on these networks and evaluate effectiveness of
intervention and vaccination protocols.  We show that medical doctors
that move between wards, but not patients or nurses each of whom is
attached to a single ward, are principal vectors (i.e. transmitters)
of pathogens.  In
accordance, we show that reallocating patients to medical doctors
reduces epidemic spreading to a larger extent than preparing single
rooms to isolate patients.  We also show that the vaccination
protocols that principally target healthcare workers, particularly
medical doctors, located in key positions determined by network
analysis are more effective than those focusing on patients or on
simply most connected individuals.

\section{Methods}

\subsection{Data}

We construct two social networks in a community hospital located in
Tokyo, based on medical records collected at two instants, namely, a
day in a weekend in September 2007 and 
a weekday in November 2007.
The use of the data in this study was approved by the Hospital Ethics
Committee.  These two datasets are different mainly in that more
nurses work on the weekday. 
Another minor difference is that
some resident doctors occupy very different parts of the
two networks because they change the department every month.
We use these two networks as two independent snapshots
and do not consider longitudinal factors
caused by, for example,
shift work, admission of patients, discharge of patients, and seasonality.

This hospital is a
482 bed tertiary referral teaching hospital, with
16 wards and 129 rooms.  Patients occupy either
shared or single rooms. There are relatively few single rooms
(37 rooms, or 7.68 \% of the total beds).
%
%
A ward has 8.06 rooms on average (max 10, min 4) and usually hosts
patients from different departments. A department
typically has patients scattered in different wards. 
Therefore, as nurses work in single wards, they are likely to
attend patients from a number of departments. 
On average, about five nurses work in a ward
(5.88 on the weekday and 4.31 on the weekend).  

\begin{figure}
\begin{center}
\includegraphics[height=7cm,width=8cm]{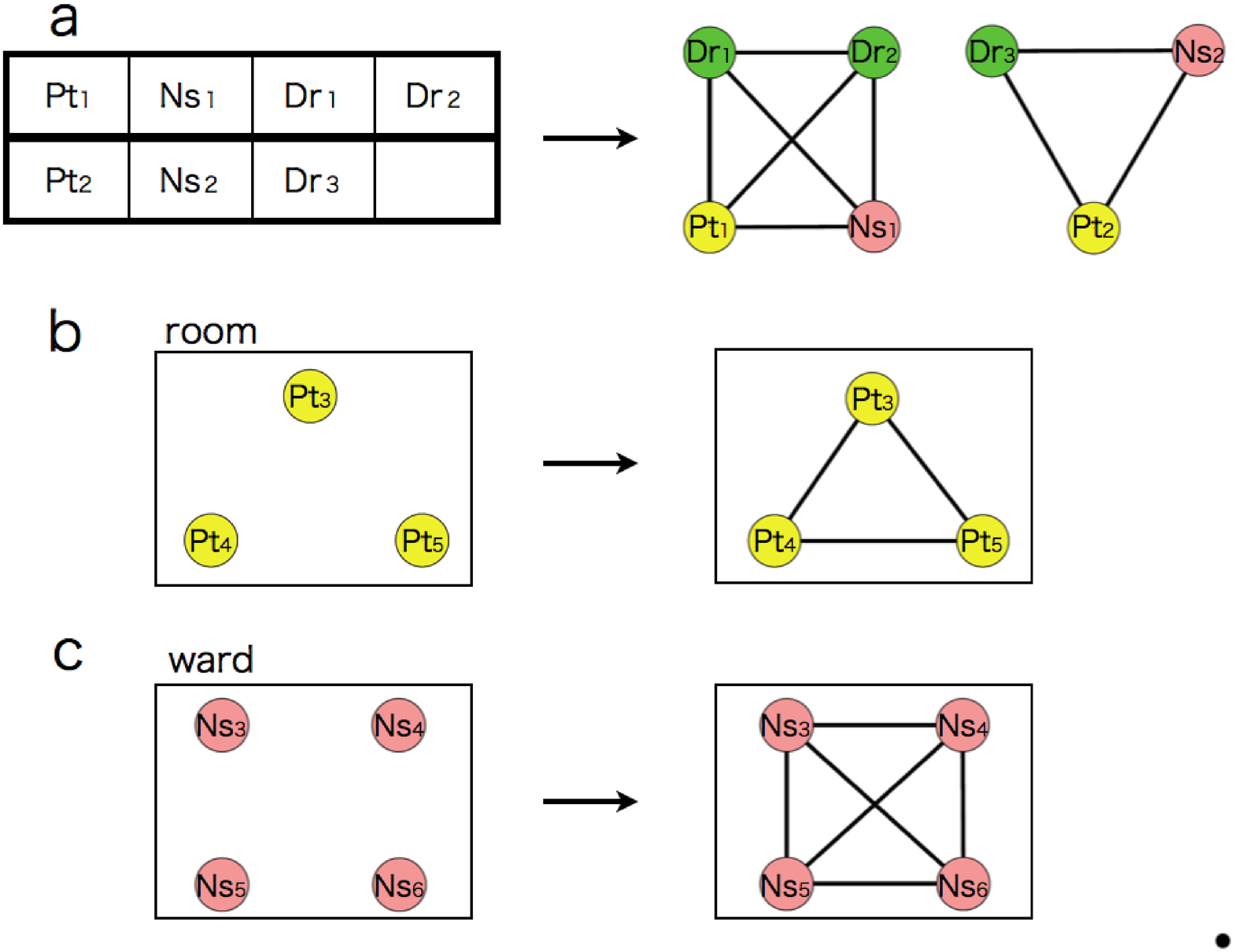}
\caption{Ueno and Masuda}
\label{fig:rule}
\end{center}
\end{figure}

As schematically shown in \FIG\ref{fig:rule}a, one line of the medical
record contains the identity of a patient accompanied by those of a
nurse and medical doctors in responsibility. A patient is assigned a
single nurse and is examined by at least one doctor. Usually, a couple
of doctors form a team so that a patient is examined by multiple
doctors in the team.  A typical team consists of two or three doctors
of different ranks. A patient is also examined by multiple doctors
when the patient registers at multiple departments, but such an
occasion is rare (none on the weekday and just one patient on the
weekend). As a note, resident doctors (25 among 123 doctors on the
weekday and 29 among 123 on the weekend) often belong to more than one
department, because they may retain their previous patients in
addition to gaining new ones every month in different departments.
All the other medical doctors are specialists and belong to only one
department. All doctors, including residents, typically visit a number
of wards over which their patients are scattered, the maximum number
of wards visited by a single doctor for these data being five.

The edges of the networks, 
which represent social contacts between pairs of individuals,
are
defined as follows. We denote patient,
nurse, and medical 
doctor, by Pt, Ns, and Dr, respectively.  The Pt-Ns, Pt-Dr,
Ns-Dr, and Dr-Dr edges (for example, a Pt-Ns edge is an edge between a
patient and a nurse) are determined based on the medical record
(\FIG\ref{fig:rule}a).  All pairs of individuals in the same line of
the medical record are connected to each other. Accordingly
doctors in the same team are connected to each other. Additionally, a
patient is connected to all the other patients in the same room
(\FIG\ref{fig:rule}b). This is the single source of Pt-Pt edges.
Similarly, a nurse is connected to all the other nurses working in the
same ward to form Ns-Ns edges (\FIG\ref{fig:rule}c).
We neglect other possible contacts, such as casual friendships
that do not appear in the medical record.

The edges of the networks are assumed to be fixed over time,
undirected, and unweighted, so that disease transmission occurs
bidirectionally with the same strength.  Even if a particular Ns-Dr
pair or Dr-Dr pair may appear in multiple lines in the medical record, we
do not double count this relation because double counting would lead
to excessive weighting of Ns-Dr and Dr-Dr edges.  For example, the two
doctors in the same team would be connected by five edges if this team
examines five patients.

\subsection{Epidemic dynamics}

We simulate the stochastic susceptible-infected-recovered (SIR) model
on the static observed networks.  Each individual takes one of the
three states represented by S (susceptible), I (infected), and R
(recovered or dead).  Initially, a prescribed index individual is
infected, and all the other individuals are susceptible. Any
susceptible individual is subject to contagion when that individual
has an infected neighbor in the social network. We assume that the
infection rate is frequency-dependent, that is, proportional to the
number of edges that an individual has \cite{Lloyd04}. We denote the
unitary infection rate by $\lambda$. Assuming that the possibility of
infection events depends the present, but not past, configuration of
S, I, and R on the network, a susceptible is infected by each infected
individual in the neighborhood with probability $1-\exp(-\lambda\Delta
t)$ in time $\Delta t$.  Note that, when $\Delta t$ is small, the
probability that a susceptible becomes infected is approximated by
$\lambda\Delta t$ times the number of infected neighbors.
Equivalently, an infected individual infects each of its susceptible
neighbors with probability $\lambda\Delta t$. The infection rate
$\lambda$ is common for different edges and individuals
unless otherwise stated.

An infected individual continues infecting neighboring individuals
until recovery occurs (and the individual enters state R) after a
random time.  The transition rate from state I to state R is set equal
to unity for normalization, so that an infected individual transits to
state R with probability $1-\exp(-\Delta t)$ for time $\Delta t$.
When $\Delta t$ is small, this probability is approximated by $\Delta
t$, and one recovery event occurs per unit time on average. This
event happens independently of the neighbors' states. Individuals in
state R are assumed to be either immune or dead. We consider both
interpretations in the following analyses. In either case, they cannot
infect, or be infected by, other individuals.

Because the transition rate from state I to state R is normalized to
unity, the infection rate $\lambda$ controls the strength of a
disease.  When $\lambda$ is large enough, with high probability, the
number of the infected grows progressively in the early stages and the
number of the susceptible decreases accordingly. After some delay,
state-R individuals begin to appear, and the number of the infected
individuals begins to decline. Finally the population is composed of
only state S and R individuals, which halts the dynamics. We
perform each numerical simulation until this situation is reached.
Because the
SIR dynamics are stochastic, we take averages over trials for
calculating the mean statistics. For each configuration, the average
is taken over 100 trials for each index case and over all the
individuals as index cases.

\section{Results}

\subsection{Network structure}\label{sub:net}

The observed weekday network contains 605 individuals (388 patients,
94 nurses, 123 medical doctors) and 3046 edges (\TAB\ref{tab:weekday}).
The characteristic
path length of the network, which is defined as the mean shortest path
length between all the pairs of individuals, $L=4.84$.  The clustering
coefficient, which counts the abundance of local connectivity by the
density of triangles normalized between 0 and 1, $C=0.534$.
A relatively small value of $L$ and a large value of
$C$ of this network
implies the small-world property \cite{Watts98}.

Although the hospital network has a considerably smaller $L$ than
spatially structured networks such as the square lattice (a regular
two-dimensional network), 
it is not perfectly mixed. If we rewire the edges
randomly with the degree (number of edges that an individual has)
of each
individual fixed, $L$ even decreases to 2.97 and
clustering is lost ($C=0.0251$).  This behavior of $L$
and $C$ upon random rewiring of edges is similar to that for the
Watts-Strogatz network model \cite{Watts98}.

The path length $L$ of the original network is larger than the
randomized network because of some localization of individuals.
First, patients are connected to each other within one room only.
Therefore, intraroom edges are dense relative to interroom edges.
Second, each patient and nurse is attached to a single ward and
only connected to patients and nurses inside that ward; only medical
doctors link different wards.  Therefore, intraward edges are dense
relative to interward edges.  Third, doctors are localized in the
sense that doctors visit only the wards with patients from their own
department and that doctors in the same team have correlated
contact patterns and mutual connections. The entire network is
composed of relatively dense subnetworks (also
called communities or network modules in network literature, e.g.
Girvan \& Newman 2002; Palla {\it et al.} 2005) 
that are loosely interlinked in a hierarchical manner.

\begin{figure}
\begin{center}
\includegraphics[height=4cm,width=4cm]{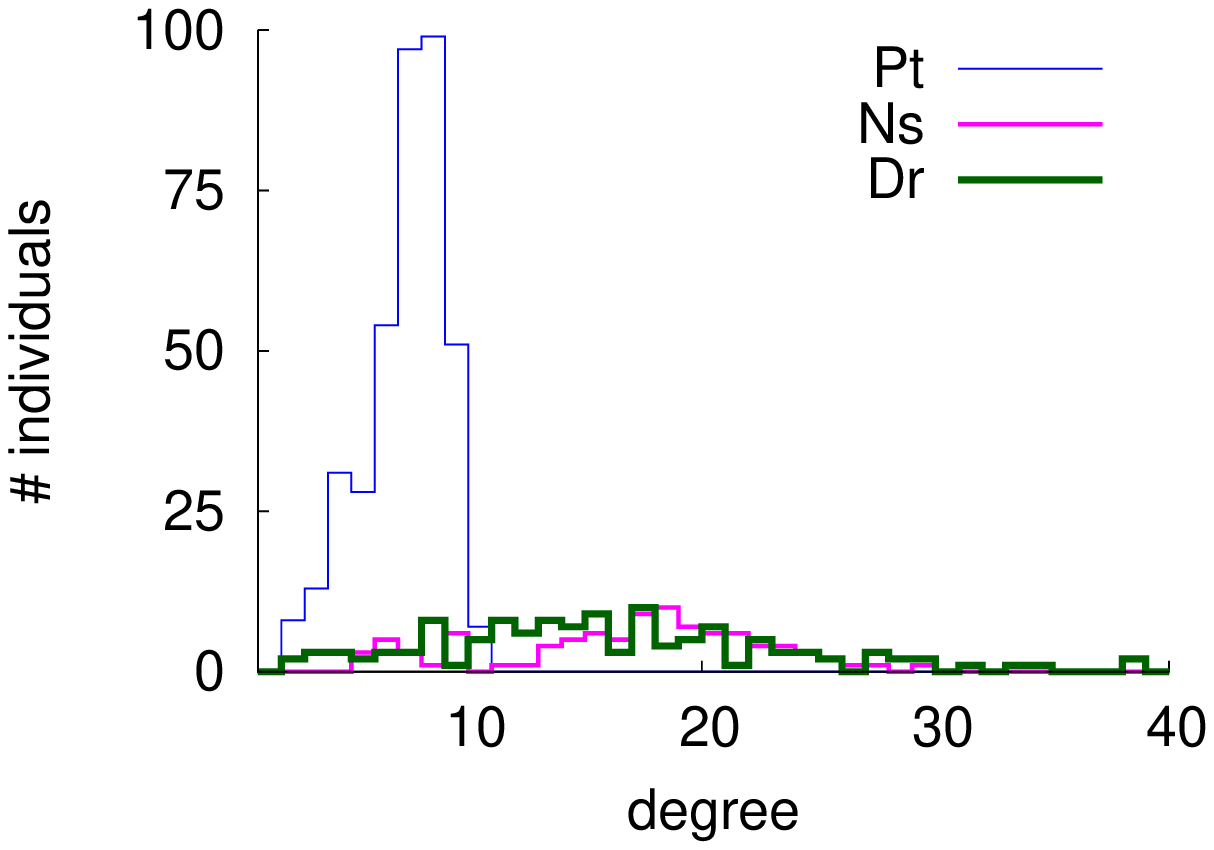}
\caption{Ueno and Masuda}
\label{fig:pk-day}
\end{center}
\end{figure}

The degree distribution is shown in
\FIG\ref{fig:pk-day} separately for each class.
Unlike social networks underlying sexually transmitted diseases
\cite{Hethcote84,Anderson86,Liljeros01} and computer viruses 
\cite{Pastor01prl},
our hospital social networks do not have right-skewed 
degree distributions. A skewed
degree distribution would imply that a fraction of caregivers with very
large degrees are extremely busy, whereas others are not.

\subsection{Basic epidemic dynamics}\label{sub:basic}

\begin{figure}
\begin{center}
\includegraphics[height=4cm,width=4cm]{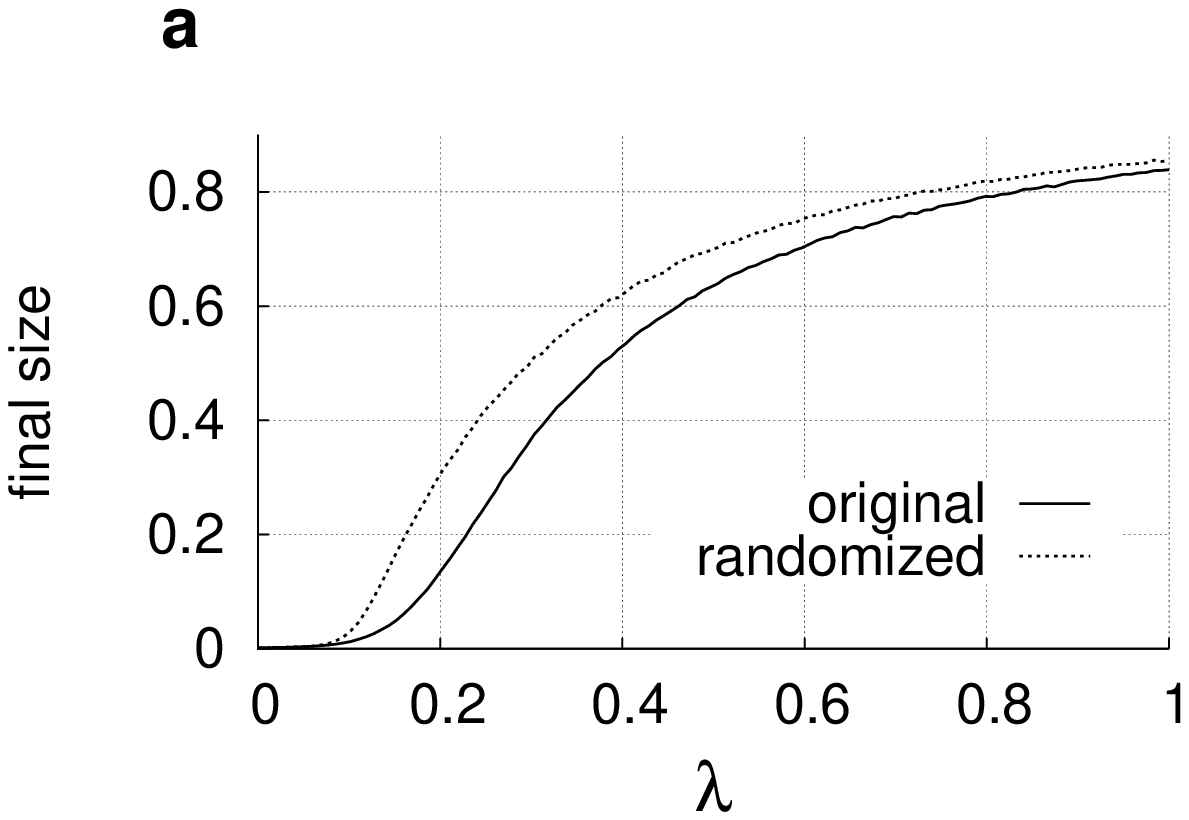}
\includegraphics[height=4cm,width=4cm]{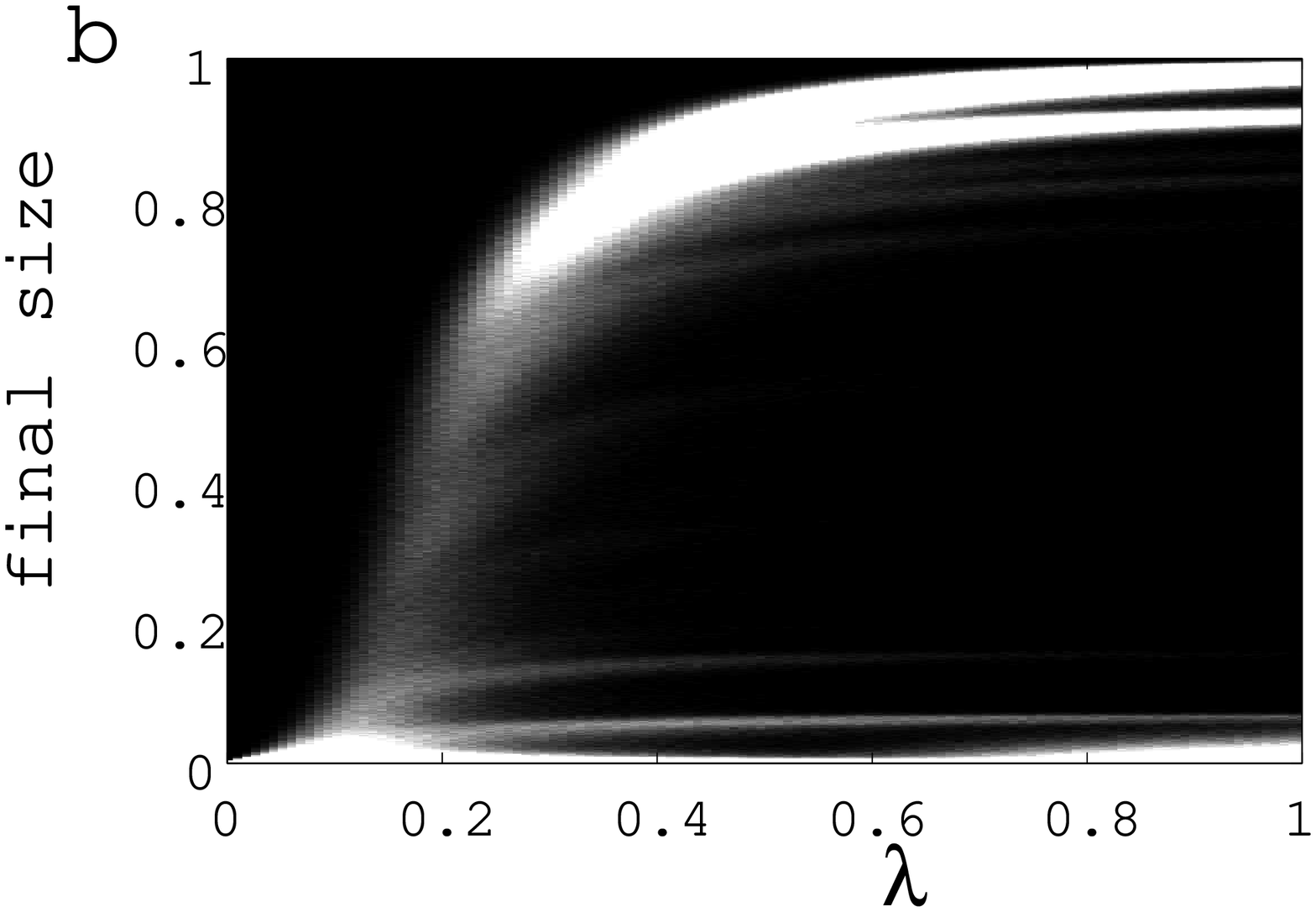}
\includegraphics[height=3cm,width=1.2cm]{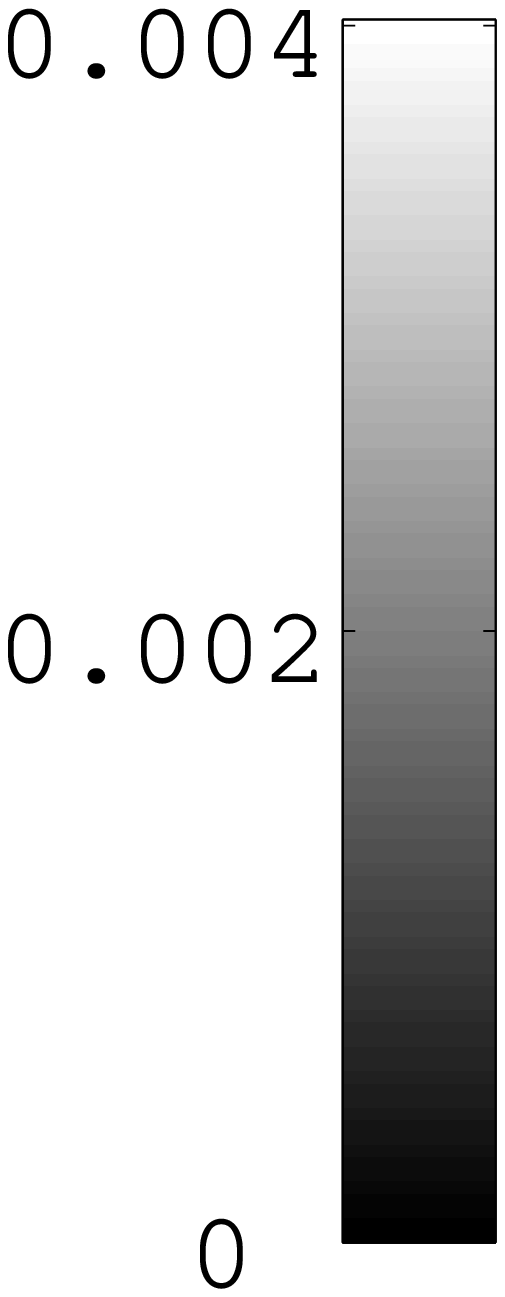}
\includegraphics[height=4cm,width=4cm]{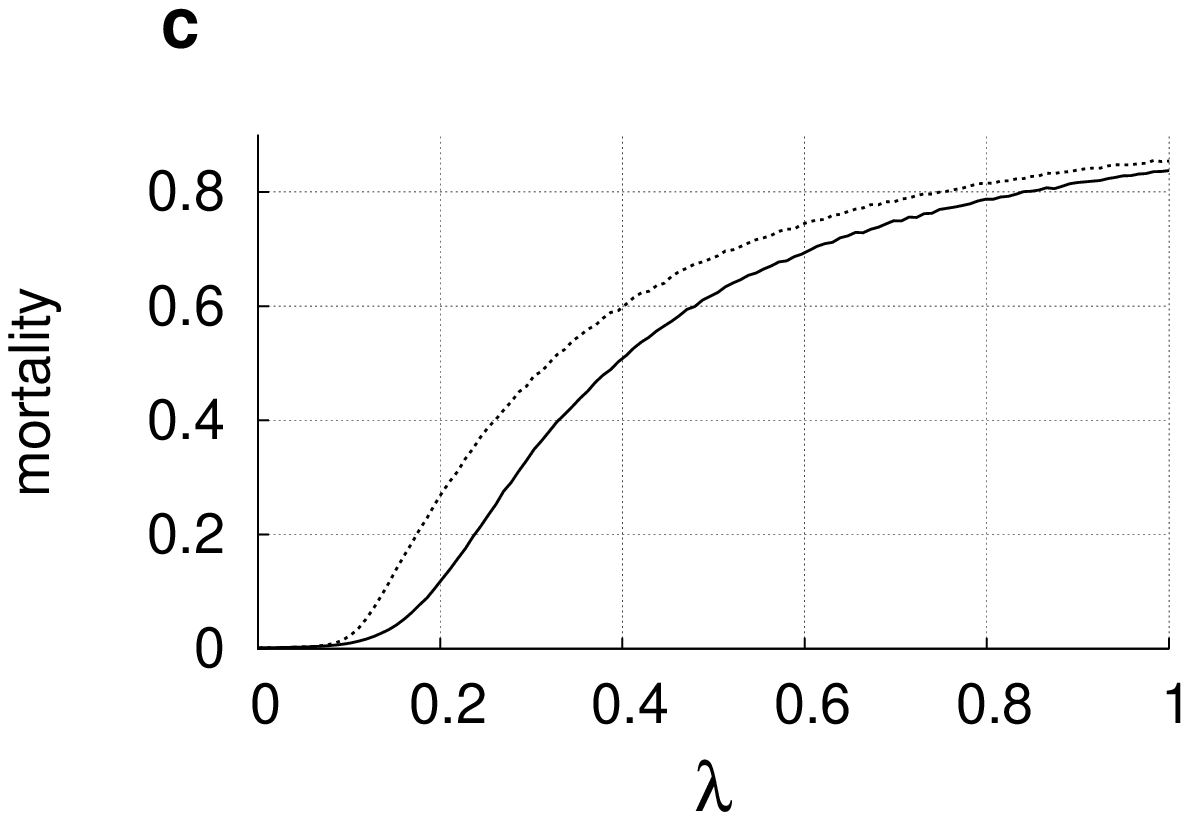}
\includegraphics[height=4cm,width=4cm]{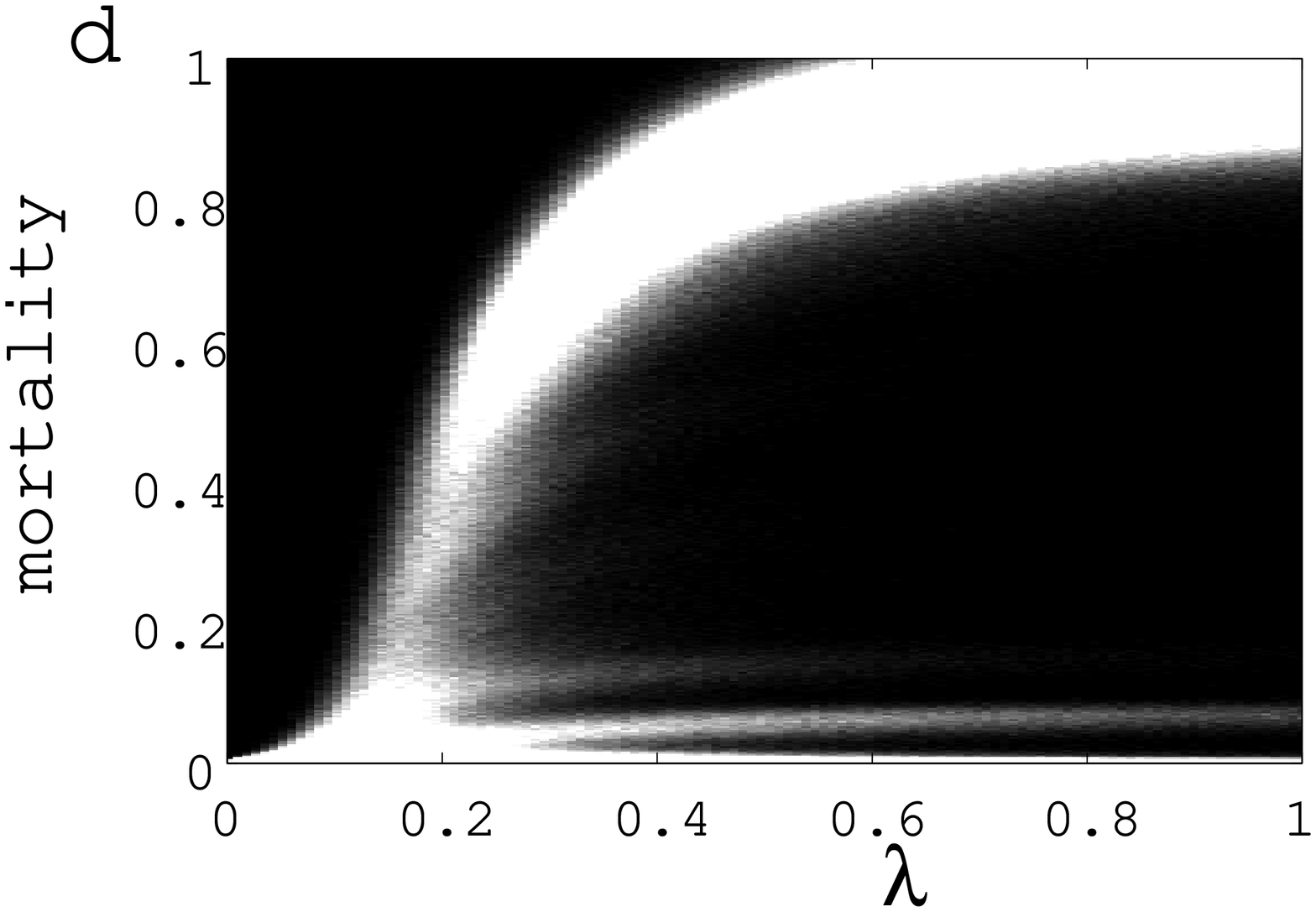}
\caption{Ueno and Masuda}
\label{fig:r-realrand-day}
\end{center}
\end{figure}

The final epidemic size is defined to be the sum of the
state R individuals regardless of the class (Pr, Ns, and Dr)
at the end of the epidemic dynamics, normalized
between 0 and 1.  The average
final size calculated numerically
is shown in \FIG\ref{fig:r-realrand-day}a (solid line)
for the weekday network.  The final size increases as the infection
rate $\lambda$ increases, taking off the zero
floor around $\lambda=0.13$, for which the basic reproduction number,
which is the expected number of secondary cases originating from an
index case, crosses unity. Beyond this point, a large-scale (i.e. major)
epidemic may occur.

The distribution of the final size is shown in
\FIG\ref{fig:r-realrand-day}b. The average of the distribution shown as
each vertical cross
section, i.e. for each value of $\lambda$, corresponds to the solid line
in \FIG\ref{fig:r-realrand-day}a. Above a threshold infection rate, the
distribution has two masses.  One thin mass is concentrated around
zero and indicates that, despite a large infection rate, the epidemic
can quickly 
decline to extinction owing to demographic stochasticity of state
I individuals (i.e. minor epidemic).  The other mass is centered
around a positive value of the final size and corresponds to the major
epidemic. Precisely speaking, the major epidemic consists of two
peaks. This is because one ward, which is devoted to the psychiatry
department, has 42 (of 605) individuals and is connected to the rest
of the network via only two junior doctors. These doctors examine
patients from departments to which they previously belonged, in
addition to patients from the psychiatry department.  Suppose that the
disease starts from the index case in this ward. Disease must pass
through the narrow channel composed of the two doctors to
spread to the rest of the population. This event occurs
with a
somewhat low probability even for very large infection rates. This causes
the lower peak of the upper mass around 0.9 in
\FIG\ref{fig:r-realrand-day}b, which is eminent for $\lambda\ge
0.8$. We confirmed that a two-mass distribution, possibly with
subdivisions of each mass reflecting modular structure of the
network, underlies each of the following numerical results.
In addition, whereas an epidemic whose final size is near
the mean value rarely occurs because of the two-mass
distribution shown in \FIG\ref{fig:r-realrand-day}b, a larger mean
final size indicates a larger probability of major epidemic.
Therefore we will show only the mean values of the
final size in later figures,

Both for nosocomial pathogens such as MRSA \cite{Forrester05} and
other pathogens such as SARS \cite{Leung04}, the case fatality of
hospitalized patients is much larger than that of healthcare
workers. The final sizes shown in 
\FIG\ref{fig:r-realrand-day}a and \ref{fig:r-realrand-day}b
represent incidences, that
is, infection events. Frequency of observing severe manifestations or
deaths needs to be discussed separately. To highlight differential
mortalities between healthcare workers and patients \cite{Leung04}, 
we assume that
only patients, not nurses or medical doctors, always die after being
infected for some time (see Discussion for justification).
The nurses and doctors in state R are assumed
to be cured after being infected.  These individuals are
counted for the final size but not for the final fraction of
mortality.  Consequently, the final fraction of mortality is defined
to be the final fraction of state R among the patients only.  The
fraction of mortality and its distribution, shown in
\FIG\ref{fig:r-realrand-day}c and \FIG\ref{fig:r-realrand-day}d,
respectively, change little from those for the final size.

The randomized network yields a
larger final size and larger mortality (dashed lines in
\FIG\ref{fig:r-realrand-day}a and \ref{fig:r-realrand-day}c)
than the original network. This is because
the randomized network has a small $L$ so that, on average,
individuals are linked by a short distance
compared to the case of the original network.

The weekend network has characteristics similar to the weekday network, 
with 521 individuals (329 patients, 69 nurses, 123 medical 
doctors), 2364 edges, $L=4.39$, and $C=0.535$ (\TAB\ref{tab:weekend}).
The randomized network has
$L=3.015$
and $C=0.0286$.
The degree distribution and the
SIR results for the original and randomized networks are shown in
\FIGS\ref{fig:pk-end} and \ref{fig:r-realrand-end}, respectively.
The final size, the mortality, and their distributions are shown in
\FIG\ref{fig:r-realrand-end}.
These results are qualitatively similar to those for
the weekday network. The following results for the intervention and
vaccination protocols are also qualitatively the same for the weekday and
the weekend network. Therefore, we explain the weekday results in the
main text and relegate the weekend results to the
Appendix.

\subsection{Interventions}

We examine the effects of three intervention methods on the
suppression of epidemic spreading. We design these intervention
methods so that they may be practiced in real situations before a
potential epidemic happens. The
cost of implementation is presumably the
smallest for method 1 and the largest for method 3.

\subsubsection{Intervention 1: reassigning patients to medical doctors}

\begin{figure}
\begin{center}
\includegraphics[height=8cm,width=8cm]{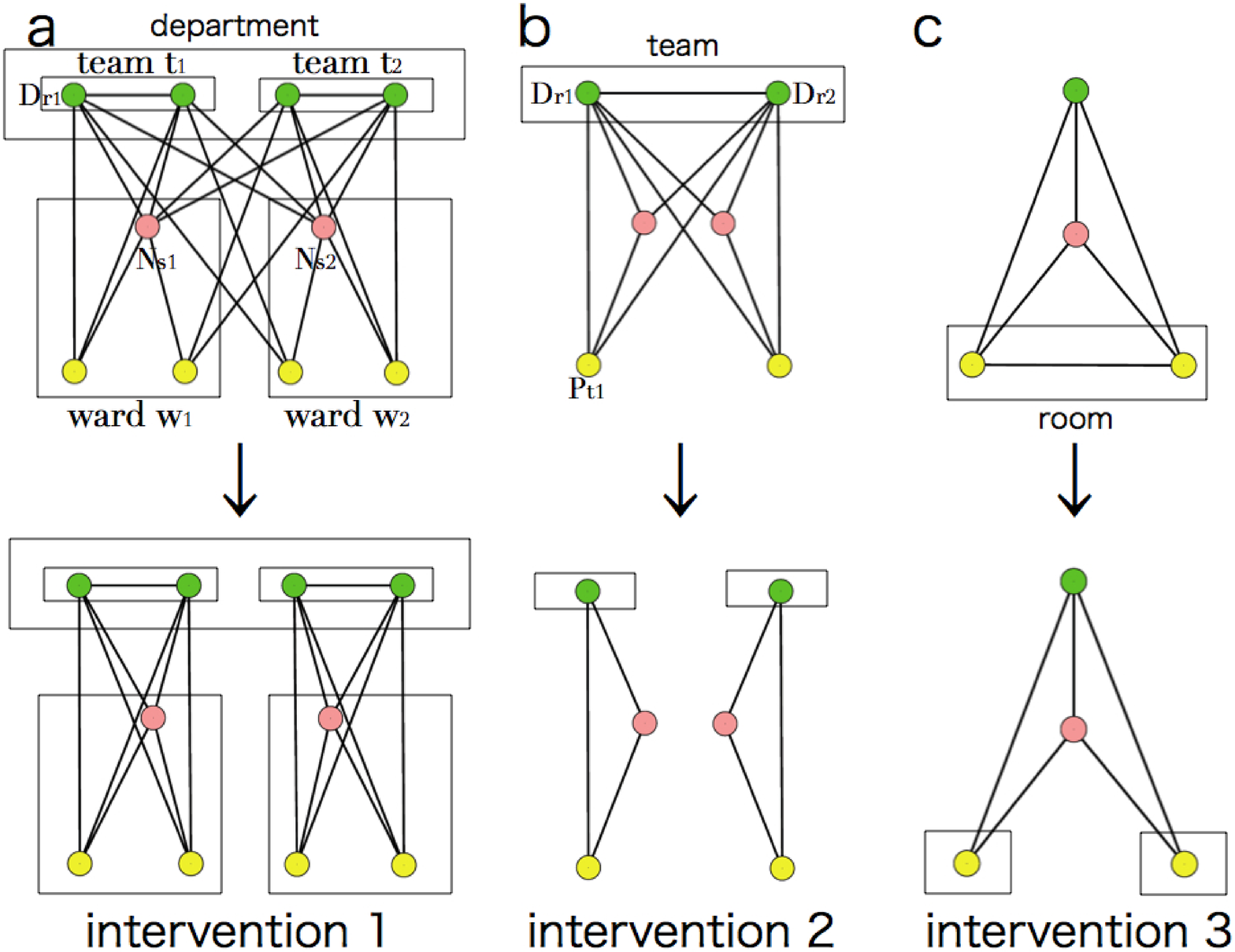}
\caption{Ueno and Masuda}
\label{fig:sch_inter}
\end{center}
\end{figure}

Medical doctors are likely to be efficient vectors
of pathogens because they
examine their patients scattered in multiple wards. In contrast,
nurses and patients are constrained to one ward.  In intervention 1,
we reduce visits of doctors to different wards by reassigning patients
to different doctors. Originally a couple of doctors in the same
department formed a team to examine a set of patients, who were
potentially scattered across different wards. Under the intervention,
we reassign the patients without destroying the team structure; we
will relax this condition in the next intervention method.  Suppose
that both teams $t_1$ and $t_2$ examine a patient in ward $w_1$
and another patient in ward $w_2$.  In the example shown in
\FIG\ref{fig:sch_inter}a, teams $t_1$ and $t_2$ trade one of their
patients.  Then team $t_1$ examines two patients in ward $w_1$ and
team $t_2$ examines two patients in ward $w_2$, thus allowing each
team to visit just one ward. We apply this operation wherever possible
so that we constrain each team of doctors to as few wards as possible,
often enabling confinement to just one ward.

By intervention 1, the numbers of Pt-Pt edges, Pt-Ns edges,
and Ns-Ns edges are preserved (\TAB\ref{tab:weekday}).
The number of Pt-Dr edges slightly decreases because
some patients are reassigned with a team of a smaller number of
doctors after intervention 
than before intervention.
The number of Ns-Dr edges decreases because a doctor visits a
smaller number of wards after the intervention.  Then, for example,
doctor Dr${}_1$ in \FIG\ref{fig:sch_inter}a tends to spend more time
with nurse Ns${}_1$, whereas this doctor is no longer connected to
Ns${}_2$. The number of Dr-Dr edges decreases because, in some departments,
some of the teams are dismissed after intervention 1. In such departments,
some doctors belong to multiple teams before intervention and to
a smaller number of teams after intervention. The edges in the original
network that link two doctors
sharing dismissed teams only
are absent in the post-intervention network.
Due to the reduction in doctor-related edges, the total number of
edges decreases from 3046 to 2827.

The numerical results for the final size and the mortality
after intervention 1 are shown in \FIG\ref{fig:r-inter-day}a and
\ref{fig:r-inter-day}b, respectively (lines with legend `1').  We find
considerable decreases in both quantities compared to
the case of the original network.  Note
that this intervention does not require introduction of additional
infrastructure.

One may argue that the suppression of epidemics comes from a
significant decrease in the number of Ns-Dr edges. To exclude this
possibility, we examine the network in which the Pt-Dr and Ns-Dr edges
are duplicated through the intervention. With this modification,
doctor Dr${}_1$ and nurse Ns${}_1$ in \FIG\ref{fig:sch_inter}a, for
example, is connected by two edges. Equivalently, we increase the
transmission rate of this single edge twofold to account for the
doubled contact rate between them, which comes from the fact that
Dr${}_1$ and Ns${}_1$ examine twice patients together (two patients
after the intervention versus one patient before the
intervention). The edge duplication is also done for a Pt-Dr edge
whenever the number of doctors that examine the patient on this edge
decreases as a result of intervention.  Consequently, the number of
Pt-Dr edges and that of Ns-Dr edges return to those of the original
network.  This implies that the quality of the medical service that we
measure by the number of Pt-Ns and Pt-Dr edges does not degrade due to
the intervention.  The number of edges in total after this modified
intervention 1 is 3016, which is close to the number for the original
network (i.e. 3046 edges; also see \TAB\ref{tab:weekday}).

The numerical results shown in \FIG\ref{fig:r-inter-day}a and
\ref{fig:r-inter-day}b (lines with legend `1 (dup)') indicate that
this modification little spoils the effectiveness of intervention 1. A
large portion of the reduced epidemics is owing to the reassignment of
patients, but not to cutting edges.

\subsubsection{Intervention 2: dissolving teams}

Next we investigate a more costly intervention method.  The team
structure of medical doctors generally enables them to reach better medical
solutions through discussion and knowledge accumulation in the team.
It also helps training junior
doctors. However, team structures provide channels for rapid disease
transmission among wards, with doctors serving as vectors. Another
related negative effect of the team structure is that patients and
nurses are subject to extra infection events because they are connected
to multiple doctors.

We dissolve the team in intervention 2. In the
schematic example shown in \FIG\ref{fig:sch_inter}b,
each patient is examined by both doctors in the team before 
intervention.
By
intervention 2, we assign each patient to one, but not two, of the
doctors. For example, Pt${}_1$ was examined by
Dr${}_1$ and Dr${}_2$ before the intervention, but only
by Dr${}_1$ after the intervention.
In practice, this intervention may be discouraged if it results in
resident or junior doctors examining patients alone.
Therefore,
we decompose the teams as much as possible under the condition that
each resident doctor remains teamed with a senior doctor.
The numbers of Pt-Pt, Pt-Ns, and Ns-Ns edges are
preserved through the intervention. 
The total number of edges drops from 3046 to 2717
owing to the reduction
in doctor-associated edges, as summarized in \TAB\ref{tab:weekday}.

The final size and the mortality are shown in
\FIG\ref{fig:r-inter-day} (lines with legend `2').
Intervention 2 is more effective than intervention 1.

To calibrate the effect of the decrease in the number of edges, we do
additional numerical simulations. As an example, suppose that
Pt${}_1$ in \FIG\ref{fig:sch_inter}b is
examined by Dr${}_1$ for twice as long
after the intervention than in the pre-intervention situation
where Pt${}_1$ is examined by two doctors.  Accordingly,
we duplicate the edge between Pt${}_1$ and Dr${}_1$. In addition, we
compensate the contact rate, or the number of edges,
between nurses and doctors by duplicating
the corresponding Ns-Dr edges, as in intervention 1. Taking the edge
duplication into account, the numbers of Pt-Dr and Ns-Dr edges, as
well as those of Pt-Pt, Pt-Ns, and Ns-Ns edges, are preserved. A
byproduct of this modification is that the service quality assessed by
the amount of the Pt-Ns and 
Pt-Dr edges is restored.  The total number of edges
after modified intervention 2 is 3004 (see \TAB\ref{tab:weekday} for
the number of each type of edges), which is close
to 3046, the number for the original network.  The persisting reduction
in the number of edges comes from that of Dr-Dr edges. As
shown in \FIG\ref{fig:r-inter-day} (lines with legend `2 (dup)'), this
modification has little influence on
the effectiveness of intervention 2. As
in intervention 1, the reduced mortality is ascribed to the
intentional changes in network structure, but not to simply cutting
edges.

\subsubsection{Intervention 3: introduction of single rooms}

In intervention 3, we attempt to prevent disease spreading among
patients by making all the rooms single, as suggested by previous
literature \cite{Cooper04bmj}.  Accordingly, we isolate the patients by
removing all the Pt-Pt edges, with the other types of edges and hence
the service quality kept intact. The schematic of this procedure is
shown in \FIG\ref{fig:sch_inter}c.  Then, the total number of edges
decreases from 3046 to 2382 (\TAB\ref{tab:weekday}).

The final size and the mortality results 
shown in \FIG\ref{fig:r-inter-day} (lines with legend `3')
suggest that intervention 3 is not as effective as interventions 1 and
2, except when the infection rate is large.
Even if we focus on the patient cases only, as 
shown in \FIG\ref{fig:r-inter-day}b,
it is more effective to reduce Dr-related edges (interventions 1 and
2) than to reduce Pt-Pt edges (intervention 3).

Note that the number of removed edges is much larger ($=642/3046$)
than for intervention 1 ($=223/3046$) and intervention 2
($=332/3046$). Generally speaking, 
cutting down more edges typically results in a
reduced epidemic size. However, this pattern does not appear
for the present network. This is
because patients are largely confined to their rooms and Pt-Pt edges do not
play an important role in epidemics on a large scale.  For a
very large infection rate, intervention 3 is better than intervention
1. However, the final size and the mortality for intervention 3
are still larger than those for intervention 2. For an extremely large
infection rate for which more than two-thirds of the population becomes
infected, intervention 3 is more efficient at control
than intervention 2.
Our claim that introducing single rooms may be ineffective
is consistent with an observation analysis
\cite{Cepeda05} and
a modeling study \cite{Bootsma06} of MRSA.

\subsubsection{Sensitivity analysis}\label{sub:sensitivity}

To show the generality of the intervention results, we perform three
sets of numerical simulations in which we change infection rates of
individuals differently. First, the rate at which each susceptible
nurse and doctor is infected is halved, representing the biological
possibility that patients are more likely to be infected than
healthcare workers.  Second, the rate at which each infected patient
infects susceptible neighbors is halved, representing the possibility
that healthcare workers have higher infection rates than patients.
Third, the rate at which each infected nurse and doctor infects
susceptible neighbors is halved, corresponding to the opposite of the
second variation.

The effects of different intervention protocols on the final size are
compared in \FIG\ref{fig:sensitivity} for the three variations
described above.  The mortality results are quantitatively very close
to the final size results in \FIG\ref{fig:sensitivity} 
(not shown). In all the three variations,
interventions 1 and 2 (particularly, intervention 2) are
more efficient than intervention 3 at low to intermediate infection
rates, which agrees with the results for the original SIR model
shown in \FIG\ref{fig:r-inter-day}.

\subsection{Vaccination}

We explore efficient vaccination strategies under the condition that
vaccines are available only to a proportion of individuals.  The
simplest strategy would be to randomly pick individuals for vaccination.
However, a number of studies suggest that
random vaccination is inferior to a
contact-based vaccination strategy in which those with largest degrees
are vaccinated
\cite{Albert00nat,Cohen03,Pourbohloul05,Bansal06}.
In theory, a degree-based vaccination strategy is effective for
well-mixed heterogeneous networks in which individuals are connected
at random.  However, as shown in \SEC\ref{sub:net}, the observed
hospital networks are not entirely mixed in this sense, but instead
have a modular, hierarchical structure.

We compare the degree-based vaccination strategy with other
strategies.  The degree is a local measure of centrality by which
individuals connected to many others are regarded to be important.
Many centrality measures have been proposed
(e.g. 
Junker {\it et al.} 2006).  In a well-mixed network, the degree is
correlated well with most other centrality measures so that hubs are
central in most senses. However, the observed hospital network is not
entirely well-mixed.  As an example, we consider the betweenness
centrality, which is a count of how often an individual mediates the
communication between arbitrarily chosen pairs of individuals along
the shortest paths \cite{Freeman79}.  Figure~\ref{fig:d-and-b}, which
compares the degree and the betweenness centrality of each individual
in the hospital, indicates that the most connected individuals are not
necessarily those with the greatest betweenness centrality values.

For a fixed centrality measure, we vaccinate the 20 individuals with
the largest centrality values, thus effectively remove them from the
network.  Then we run the SIR dynamics on the reduced network.  The
drops in the final size and the mortality size relative to those for
the original network are interpreted as the effectiveness of a
vaccination method.

The final size (\FIG\ref{fig:vac-day}a) and the mortality
(\FIG\ref{fig:vac-day}b) with each of
four vaccination protocols are compared to those
for the original network. Degree-based vaccination is more effective
at reducing epidemics than random vaccination (compare lines with
legend `random' and ones with legend `degree' in
\FIG\ref{fig:vac-day}), which agrees with the results in previous
literature \cite{Albert00nat,Cohen03,Pourbohloul05,Bansal06}. The
degree-based vaccination is superceded by the betweenness-based
vaccination for a wide range of $\lambda$, agreeing with the previous
modeling work
by Holme {\it et al.} (2002).
The betweenness-based
vaccination is effective because interward communication is inhibited
after vaccination of key individuals. These key individuals are mostly
medical doctors and sometimes nurses. In our simulations, the two doctors
who connect the nearly isolated ward to the remainder of the network,
whom we mentioned in
\SEC\ref{sub:basic},  have
the largest and the third largest betweenness
centrality values. These doctors
are actually vaccinated in the betweenness-based protocol.  A disease does
not enter a subnetwork if the entire network is fragmented after
removing such key individuals.  Even if the network is 
not fragmented after
vaccination, a chain of transmission has to make detours, which makes
disease spreading less likely.  This is presumably
because important shortcuts
owned by central individuals have been cut out.

If we recalculate the betweenness centrality every time
after removing the most
central individual, the network generally falls apart
in an earlier stage
\cite{Girvan02,Holme02pre_attack}. By doing so, epidemic
spreads are more suppressed (\FIG\ref{fig:vac-day}a and
\ref{fig:vac-day}b, lines with legend `recal-betw').  The improvement
is eminent for a large infection rate for which the disease
tends to spread over many wards, infecting more than half 
the individuals in the hospital ($\lambda$ around 0.8 and beyond).
However, the final size and the
mortality for nearly threshold infection rates ($\lambda$ around 0.2)
are slightly increased compared to the standard betweenness-based
vaccination.

Our vaccination protocols up to this point do not take into account
whether vaccinated individuals are patients, nurses, or medical
doctors. We compare these class-mixed protocols to ones
in which central individuals of a specific class are
vaccinated with priority.  For example, we vaccinate 20 doctors with
the largest centrality values (doctor-first vaccination).

Based on the betweenness centrality, the final size 
and the mortality
for patient-first, nurse-first, doctor-first, and class-mixed
vaccination protocols are shown in \FIG\ref{fig:vac-day}c and
\ref{fig:vac-day}d. The corresponding results based on
the recalculated betweenness centrality are shown in
\FIG\ref{fig:vac-day}e and \ref{fig:vac-day}f.  For both centrality
measures, the nurse-first strategy is less effective than the
class-mixed and the doctor-first strategies. This is because, even though
some nurses have large centrality values 
(\FIG\ref{fig:d-and-b}b)
comparable to those of the
most central doctors (\FIG\ref{fig:d-and-b}c), vaccinating nurses does
not lead to dissociation of different wards; only doctors
connect wards. The patient-first strategy is the least effective
at reducing both the final size 
(\FIG\ref{fig:vac-day}c and \FIG\ref{fig:vac-day}e) and
the mortality (\FIG\ref{fig:vac-day}d and \FIG\ref{fig:vac-day}f),
because the patients have low centrality values irrespective of
centrality measure
(\FIG\ref{fig:d-and-b}a). The doctor-first
strategy is more effective
than class-mixed strategy for the non-recalculated
betweenness centrality (\FIG\ref{fig:vac-day}c and \ref{fig:vac-day}d)
and vice versa for the recalculated betweenness centrality
(\FIG\ref{fig:vac-day}e and \ref{fig:vac-day}f).  The doctor-first
strategy is better for both the non-recalculated and recalculated
betweenness centrality for the weekend data (\FIG\ref{fig:vac-end}c-f).

We have also examined the vaccination protocols based on
other centrality measures.
For each of the centrality measures examined (closeness centrality,
eigenvector centrality, power centrality,
information centrality, and random-walk centrality;
see 
Junker {\it et al.} 2006, for a list of centrality measures),
the overall tendency that
the class-mixed and doctor-first vaccinations
are superior to the nurse-first vaccination, which supercedes the
patient-first vaccination, is reproduced (not shown).
Betweenness-type centrality
measures such as the random-walk centrality turn out to be
more effective at reducing epidemics than
others. Whether the class-mixed strategy is
superior to the doctor-based strategy depends on the
centrality measure and whether the weekday network or the weekend
network is used.

\section{Discussion}

We have investigated the effects of intervention and vaccination
strategies using numerical simulations of the SIR model on hospital
social networks. Both the intervention and vaccination results
suggest that
medical doctors 
that link different wards are main spreaders of nosocomial
pathogens.  To rewire the Pt-Dr relationship (interventions 1 and 2;
particularly, intervention 2)
is more efficient than suppressing Pt-Pt communication (intervention
3).  Note that, in spite of the relative ineffectiveness of
intervention 3, it involves construction of single rooms, which
presumably costs more than interventions 1 and 2, and may result in
reduction in the hospital admission capacity.  Vaccinating doctors
first suppresses epidemics more than vaccinating patients or nurses
first. Importantly, doctor-centered containment strategies are more
effective than patient-centered ones at reducing not only the final size
of the entire population but also that of the patients.

We have found that the observed social networks do not have
extremely right-skewed distributions 
of the degree, which differs from
many other
disease-related networks that
have scale-free degree distributions
(e.g. 
Albert {\it et al.} 2000; Liljeros {\it et al.} 2001).  Instead, our
observed networks in the hospital have modular and hierarchical
structure regulated by wards, rooms, teams of medical doctors, and
departments. Different modules are bridged by healthcare workers,
especially by medical doctors.
Therefore, the common result that scale-free networks boost epidemics
owing to hubs, which hold for well-mixed networks
with right-skewed degree distributions
\cite{Hethcote84,Anderson86,Albert00nat,Pastor01prl,Meyers03,Christley05,Meyers05},
seems to be rather
irrelevant to the current situation, as has been pointed out
for other situations \cite{Holme02pre_attack,Watts05,Green06,Ichinomiya}.
For our networks,
we have shown that the vaccination protocols based on the betweenness
centrality, which target individuals that tend to maintain shortcuts,
are more effective than that based on the degree.  The
network structure identified in this work may serve as a basis for
network-based quantification studies.

Healthcare workers have been suggested to be primary vectors and
spreaders because they visit different wards
\cite{Elder96,Carman00,Meyers03,Ho03,CDR05,Burls06}.  Therefore,
vaccinating healthcare workers may be more effective than vaccinating
patients in preventing nosocomial infection \cite{Carman00,Meyers03}.
These results are consistent with ours. We have further distinguished
roles of different classes of healthcare workers, namely, medical
doctors and nurses, which was ignored in an observation study of
nosocomial infection in which only patients were considered
\cite{Liljeros07}. We also examined the effects of individuality in
terms of their position in the networks. We suggest that healthcare
workers, doctors in particular, with large betweenness centrality
values are of high mobidity in nosocomial infection.  Because the
betweenness centrality is an ad hoc measure for our purpose, there may
be alternative measures more directly related to mobidity of
individuals.

An implicit assumption underlying the vaccination strategies based on
the betweenness and most other centrality measures is that the
information on the global network structure is available.  This
assumption is not justified for urban community infections where the number
of individuals involved is large and exhaustive contact tracing is
infeasible.
In this situation, a powerful vaccination strategy based on
only the local information is ring vaccination in which
vaccinations are provided along chains of contacts
reconstructed by querying friends' names
\cite{Ferguson01,Cohen03,Pourbohloul05}. An alternative
strategy is to abandon
heterogeneous contact rates at the level of individuals
and employ a population-based approach in which
distributions of contact rates are approximated using discrete
population groups such as age groups
\cite{Wallinga06}.  
However, the use of global
information can be practical in the nosocomial setting because healthcare 
facilities are limited in size. In
addition, the information about contacts between individuals may be more
available than in the case of urban community infection.  We have obtained
this information by analyzing medical records.

To measure the mortality, we have assumed that the patients always die
after being infected, whereas healthcare workers do not.  However,
the results are robust with respect to the definition of
mortality. The difference between (i) the pattern of infection events
for the entire population (\FIGS\ref{fig:r-realrand-day}a,
\ref{fig:r-realrand-day}b, \ref{fig:r-inter-day}a, and
\ref{fig:vac-day}a, \ref{fig:vac-day}c, \ref{fig:vac-day}e) and (ii)
that for the patients (\FIGS\ref{fig:r-realrand-day}c,
\ref{fig:r-realrand-day}d, \ref{fig:r-inter-day}b, and
\ref{fig:vac-day}b, \ref{fig:vac-day}d, \ref{fig:vac-day}f) is
unnoticeable, even quantitatively. This implies that, even if all the
infected individuals are assumed to eventually die (corresponding to
(i)), or healthcare workers die of infection with an intermediate
probability (corresponding to mixture of (i) and (ii)), the results
would rarely change. In addition, the results are expected to
be robust against reduction in the case fatality of patients because 
reinterpretation of a fraction of state R
patients as being recovered instead of dead little changes
the results.

There are some limitations to the current work. First, we have
neglected the weight of edges, or equivalently, heterogeneity in the
transmission rate of individuals, except partial consideration in
interventions 1 and 2 and in the sensitivity analysis. In reality,
a Pt-Pt edge may be stronger than a Pt-Dr edge because
the patients in the same room spend much time together, whereas a
doctor examines a patient possibly for a couple of minutes per day.
We have ignored this factor because the information available is limited
(but see
Bansal {\it et al.} 2006) and because the actual edge weights will
depend on the type of diseases. Related to this, we have assumed
frequency-dependent transmission in which the infection rate is
proportional to the number of edges involved. Although
this assumption is valid for sexually transmitted diseases \cite{Lloyd04},
whether this assumption holds for
nosocomial pathogens is not known.

Second, our networks do not include some types of relations and
individuals. For example, there are three medical doctors for diabetes
and one doctor for infection control in the hospital.
These specialist 
doctors examine patients belonging to different departments
upon request. However, our medical records miss such Pt-Dr
relations.  These doctors are expected to interact with more wards
than ordinary doctors to play a leading role in epidemic spreading.
We have also neglected radiographers and pharmacists that interact
with patients in multiple departments, and visitors that would link
the hospital to the outside community.
We have also omitted social contacts outside of 
official hospital business. For example, 
two doctors may be friends such that they have lunch together
on a daily basis, and
doctors in different teams may share a office.

Third, we have assumed the simple transmission rule defined by the SIR
model. Nosocomial pathogens such as MRSA and VRE cause asymptomatic
infection in which carriers, of which there are many types, do not
necessarily infect others. In the SIR model, incubation periods are
ignored and all the individuals are as susceptible and
infectious as others at a common infection rate. However, given the
robustness of our results against heterogeneous infection 
rates for different individuals shown in \SEC\ref{sub:sensitivity},
we expect that our finding is extended to more general situations of
disease spreading.
Modeling dynamics of
particular nosocomial pathogens is an important
future problem.

Fourth, we have not considered dynamics of social networks
\cite{Volz07}, except that we have confirmed that the
main results are qualitatively the same at two
time points (see Appendix for the other data set).
In reality, the members of a hospital change
on a very short time scale due to, for example,
admission and discharge of patients \cite{Bootsma06,Liljeros07}

In conclusion, the present study examined the spread of nosocomial
infection in explicitly structured and observation-based social
networks in a hospital, emphasizing the importance of controlling
medical doctors as a potential major vector of diseases. Despite the
necessity of various improvements, we believe that our exercise
highlights specific hierarchical structure of social networks of
hospitals and associated disease spreading patterns.

\section*{Acknowledgments}
We thank Hiroshi Nishiura and
Takashi Ichinomiya for critical reading of the manuscript and
Yuki Yasuda for helpful comments on this work.
NM is supported by Grants-in-Aid for Scientific Research from MEXT,
Japan (Nos. 20760258 and 20540382).

\section*{Appendix: results for the weekend network}

In the main text, we explained the numerical results for the weekday
network data. We show the
results for the weekend network data
in \FIGS\ref{fig:pk-end}, \ref{fig:r-realrand-end},
\ref{fig:r-inter-end}, \ref{fig:vac-end}, and \TAB\ref{tab:weekend}.
The main results are
qualitatively the same at these two time points.  Some differences are
in the relative effectiveness of the class-based vaccination
strategies (\FIG\ref{fig:vac-end}c-f).

\newpage

\pagestyle{empty}

Figure and Table captions

\bigskip

Figure 1: Construction of
(a) Pt-Ns, Pt-Dr, Ns-Dr, Dr-Dr edges, (b) Pt-Pt edges, and (c)
Ns-Ns edges.

\bigskip

Figure 2: Degree distributions of the weekday network for different
classes of individuals.

\bigskip

Figure 3: Disease spread in the weekday network.  (a) Final size for
the original network (solid lines) and the randomized network (dashed
lines).  (b) Distribution of the final size.  (c) Fraction of
mortality. (d) Distribution of the fraction of mortality. A brighter
point in (b) and (d) corresponds to a larger probability density.

\bigskip

Figure 4: Schematics of three intervention methods.

\bigskip

Figure 5: Effects of different intervention protocols for the weekday
network. (a) Final size and (b) fraction of mortality. Different
lines correspond to the original observed network (`original'), networks
after intervention 1 (`1'), intervention 1 with edge
duplication (`1 (dup)'), intervention 2 (`2'),
intervention 2 with edge duplication (`2 (dup)'), and
intervention 3 (`3').

\bigskip

\begin{figure}
\begin{center}
\includegraphics[height=4cm,width=4cm]{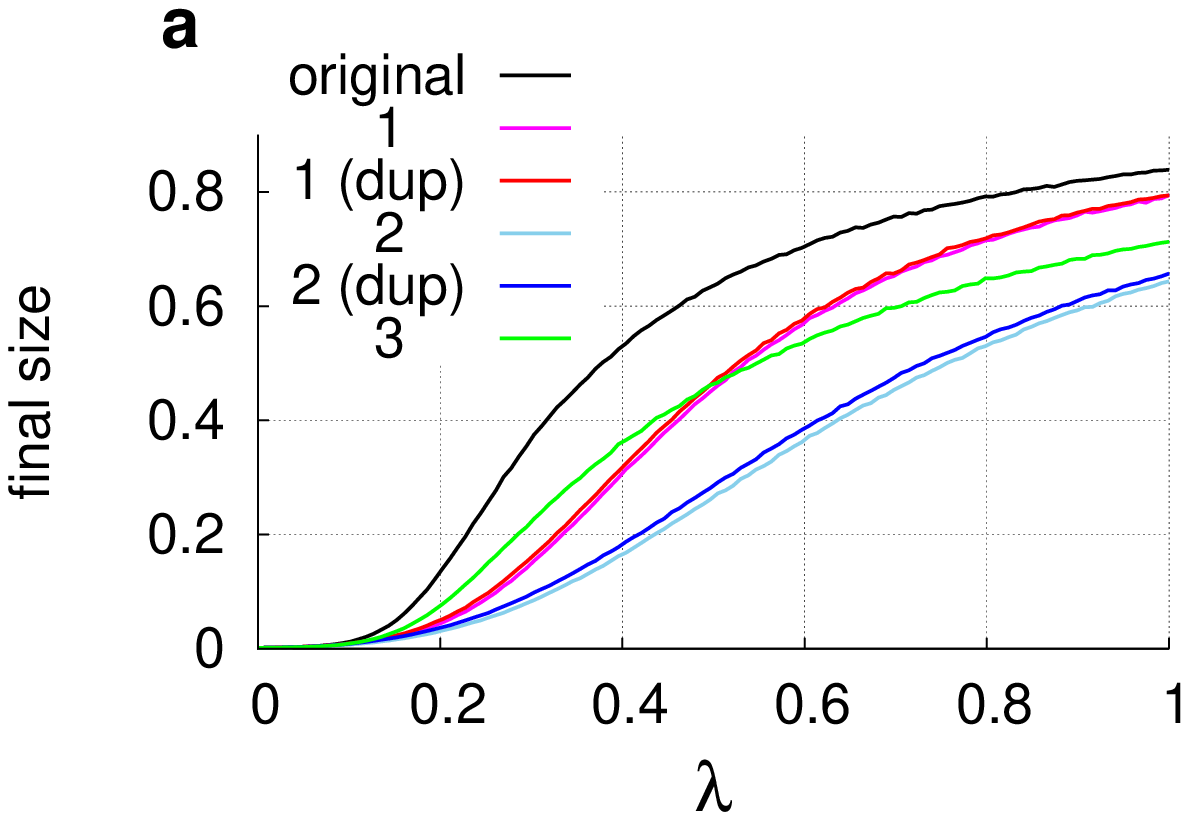}
\includegraphics[height=4cm,width=4cm]{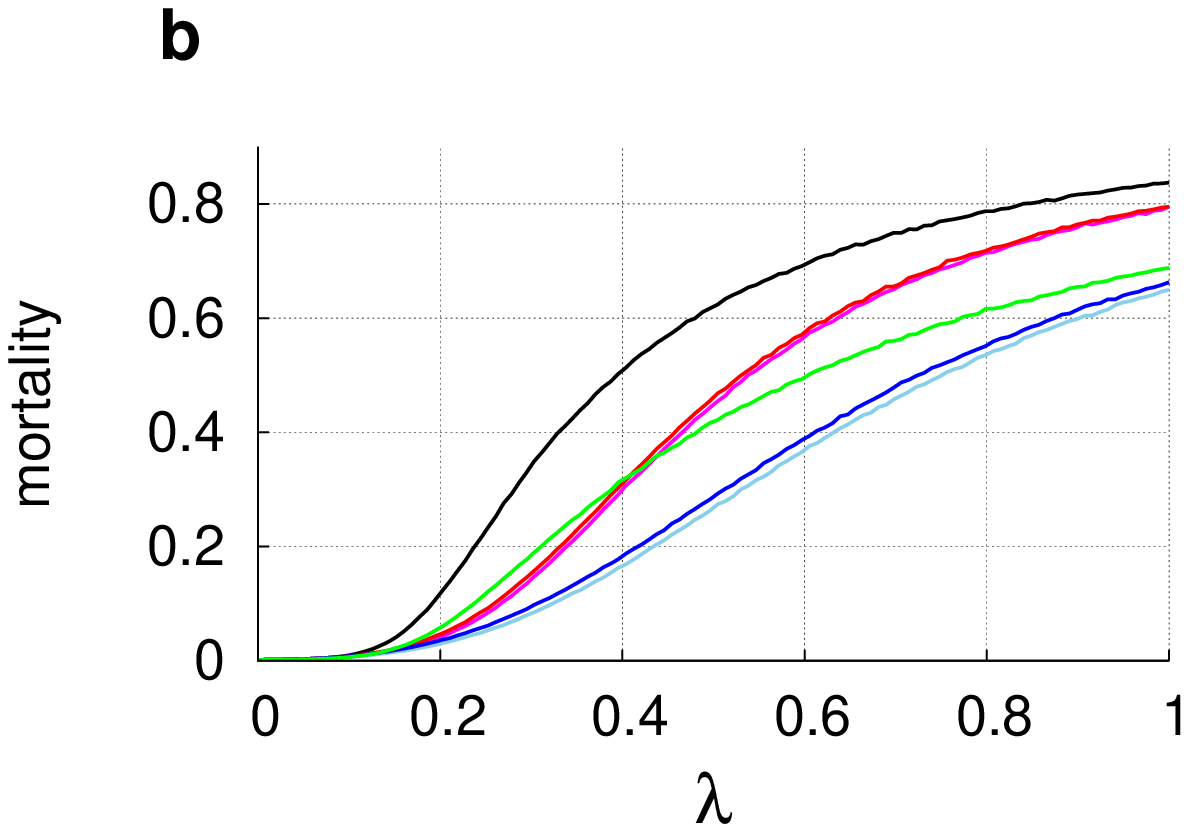}
\caption{Ueno and Masuda}
\label{fig:r-inter-day}
\end{center}
\end{figure}

\begin{figure}
\begin{center}
\includegraphics[height=4cm,width=4cm]{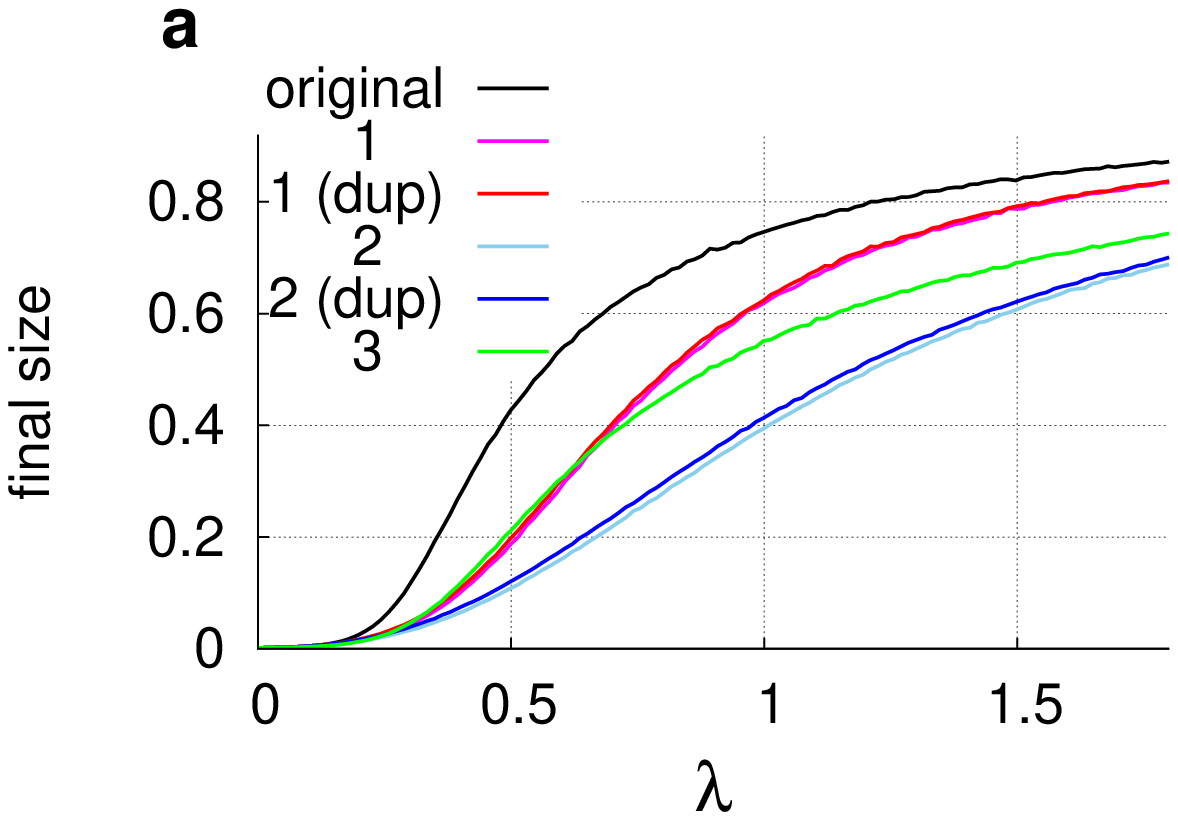}
\includegraphics[height=4cm,width=4cm]{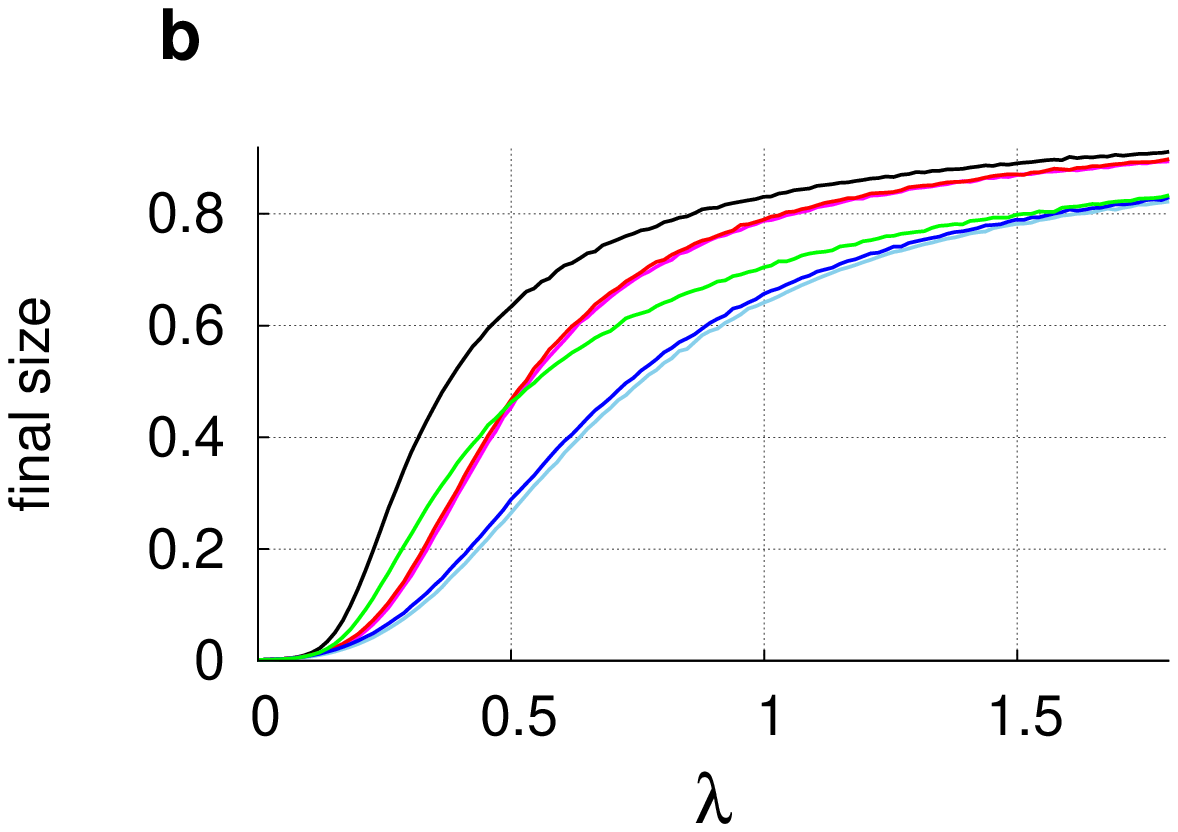}
\includegraphics[height=4cm,width=4cm]{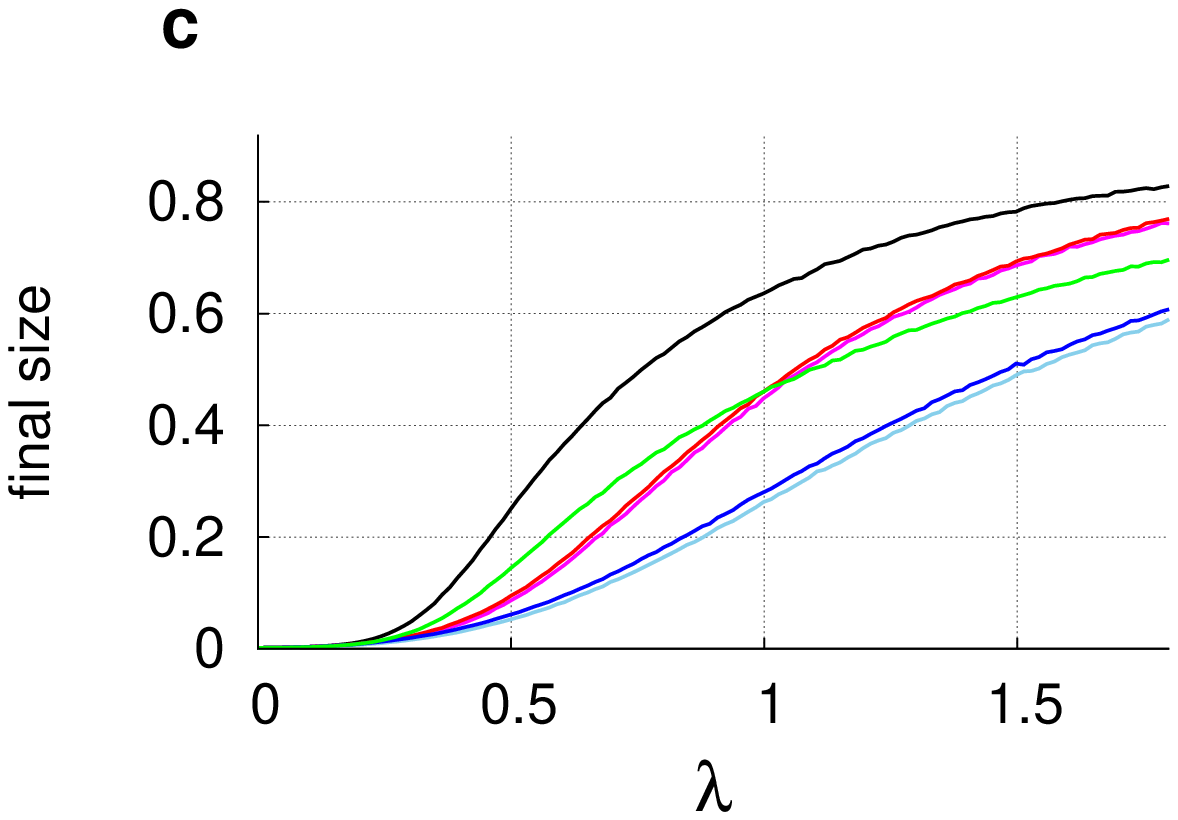}
\caption{Ueno and Masuda}
\label{fig:sensitivity}
\end{center}
\end{figure}

\begin{figure}
\begin{center}
\includegraphics[height=4cm,width=4cm]{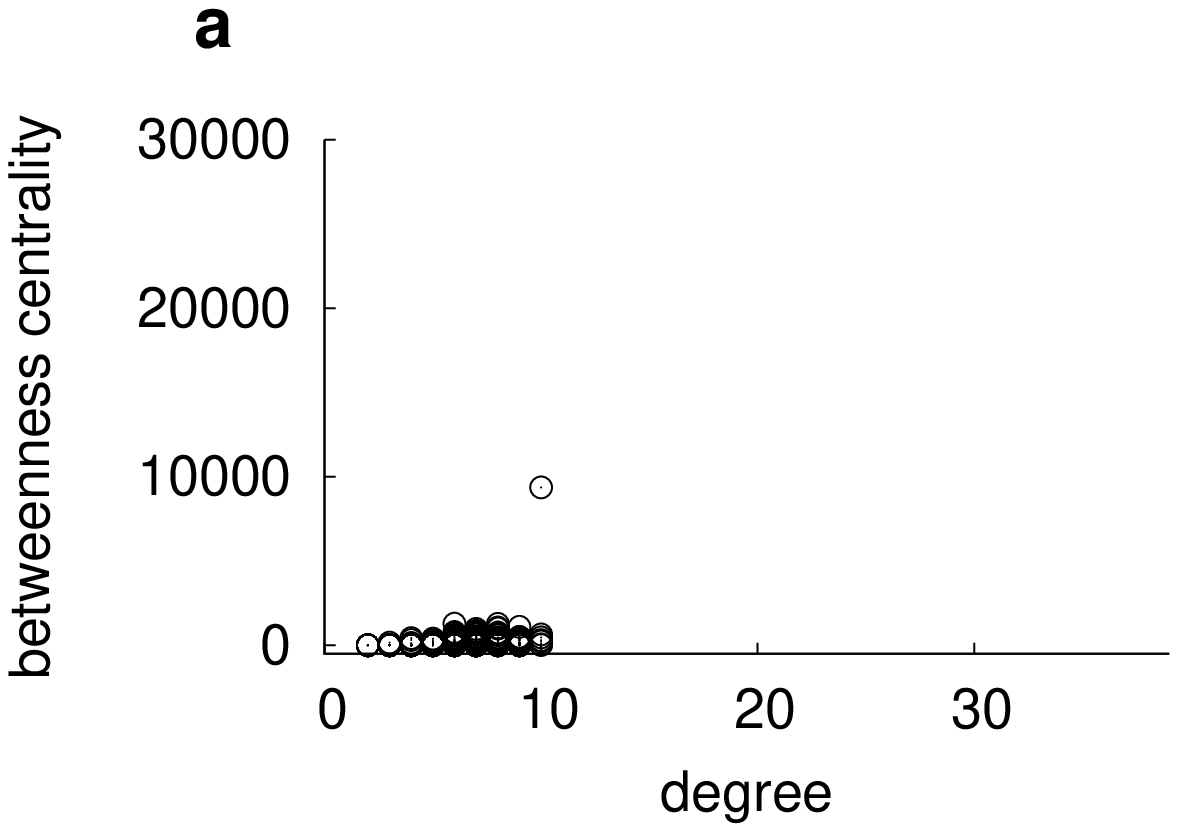}
\includegraphics[height=4cm,width=4cm]{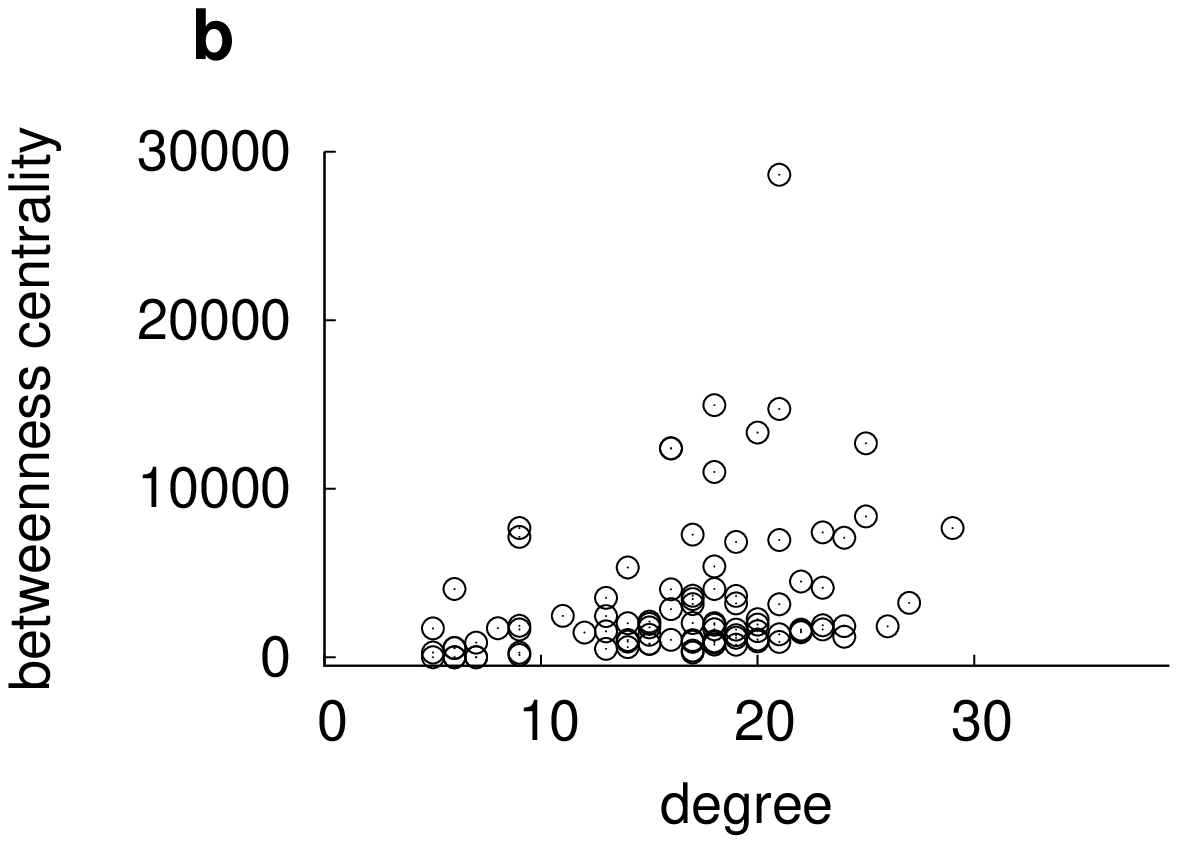}
\includegraphics[height=4cm,width=4cm]{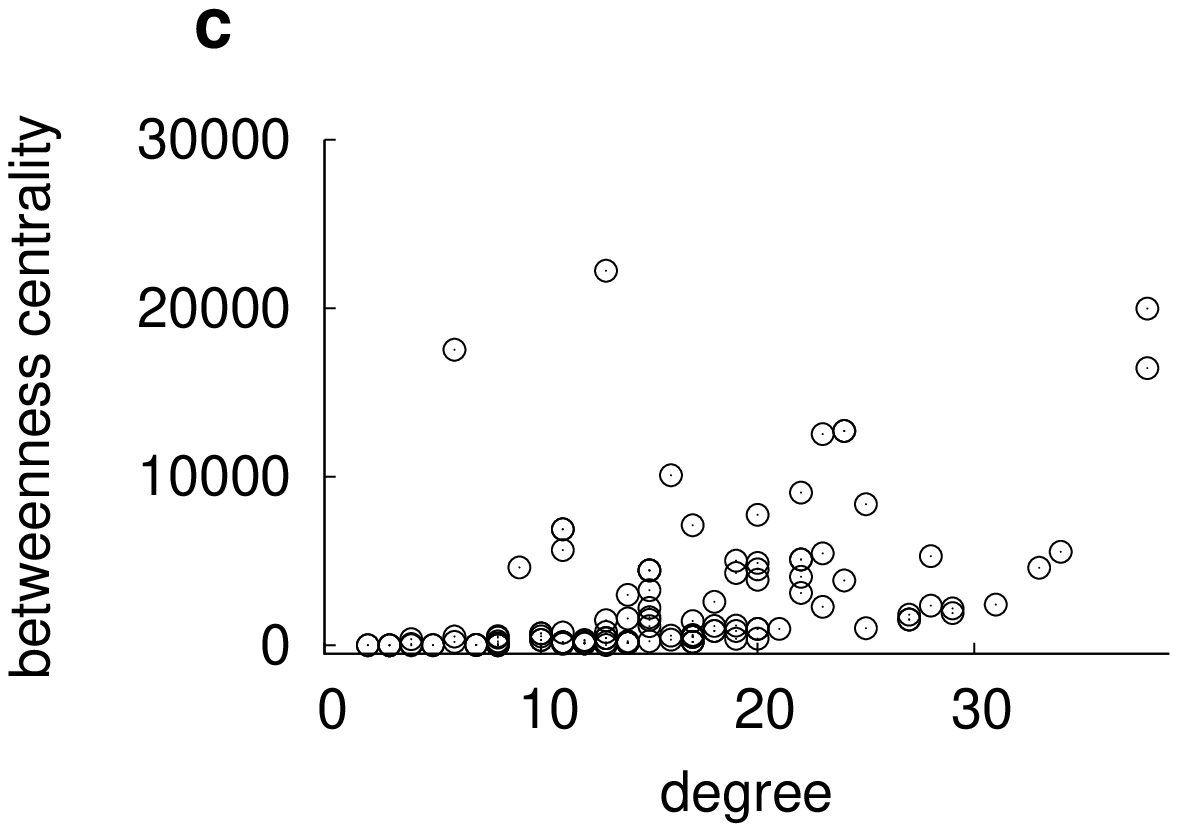}
\caption{Ueno and Masuda}
\label{fig:d-and-b}
\end{center}
\end{figure}

\begin{figure}
\begin{center}
\includegraphics[height=4cm,width=4cm]{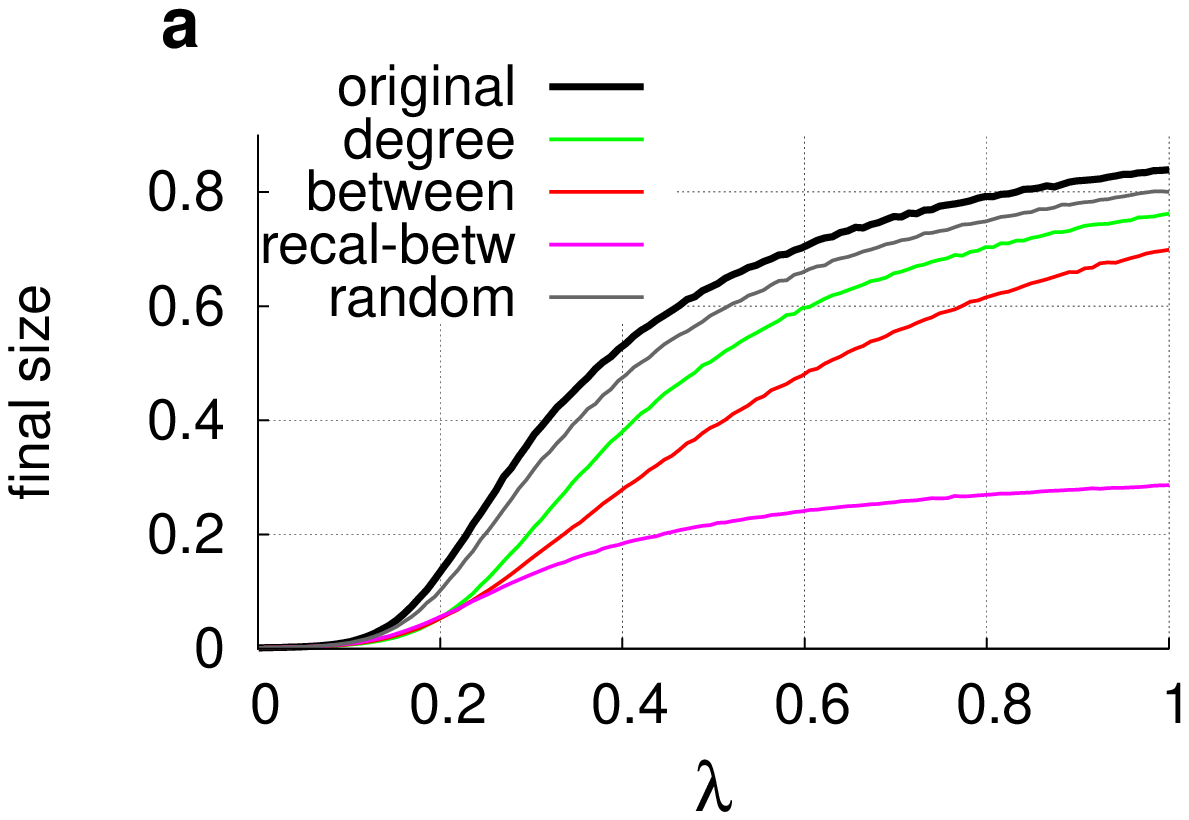}
\includegraphics[height=4cm,width=4cm]{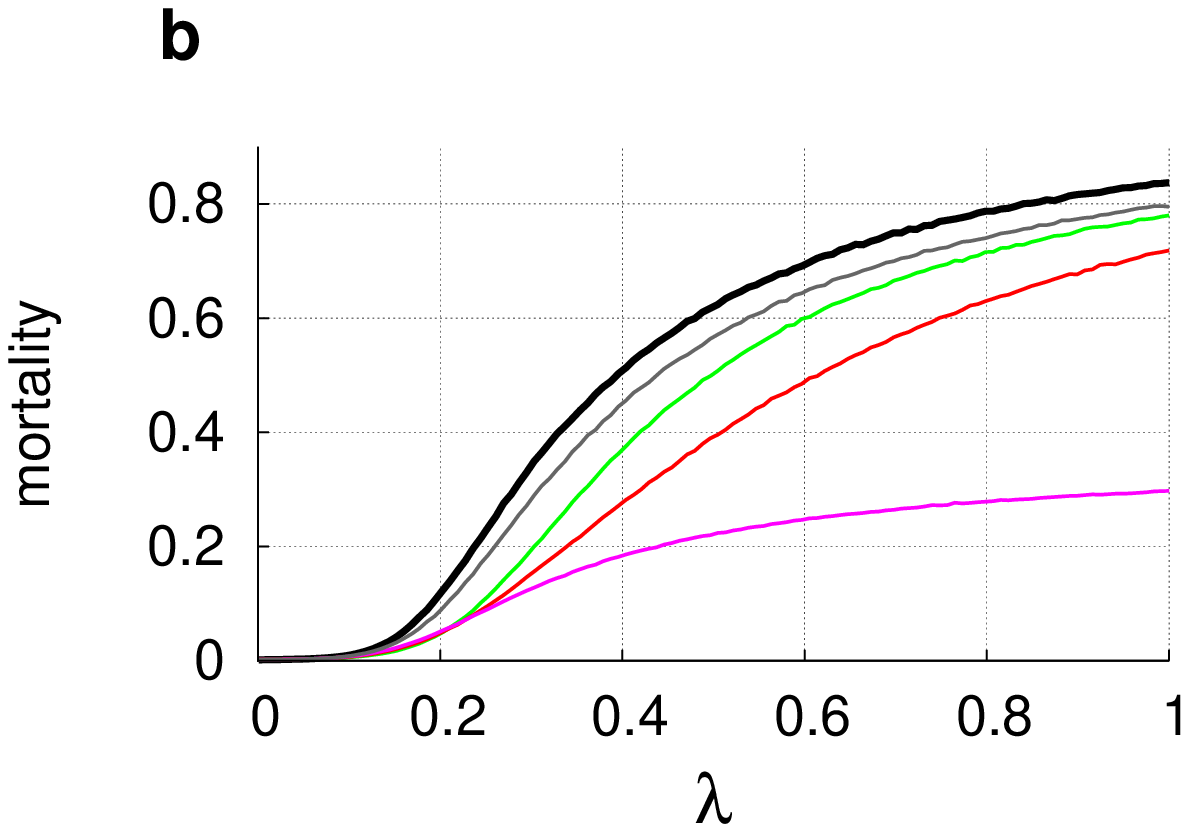}
\includegraphics[height=4cm,width=4cm]{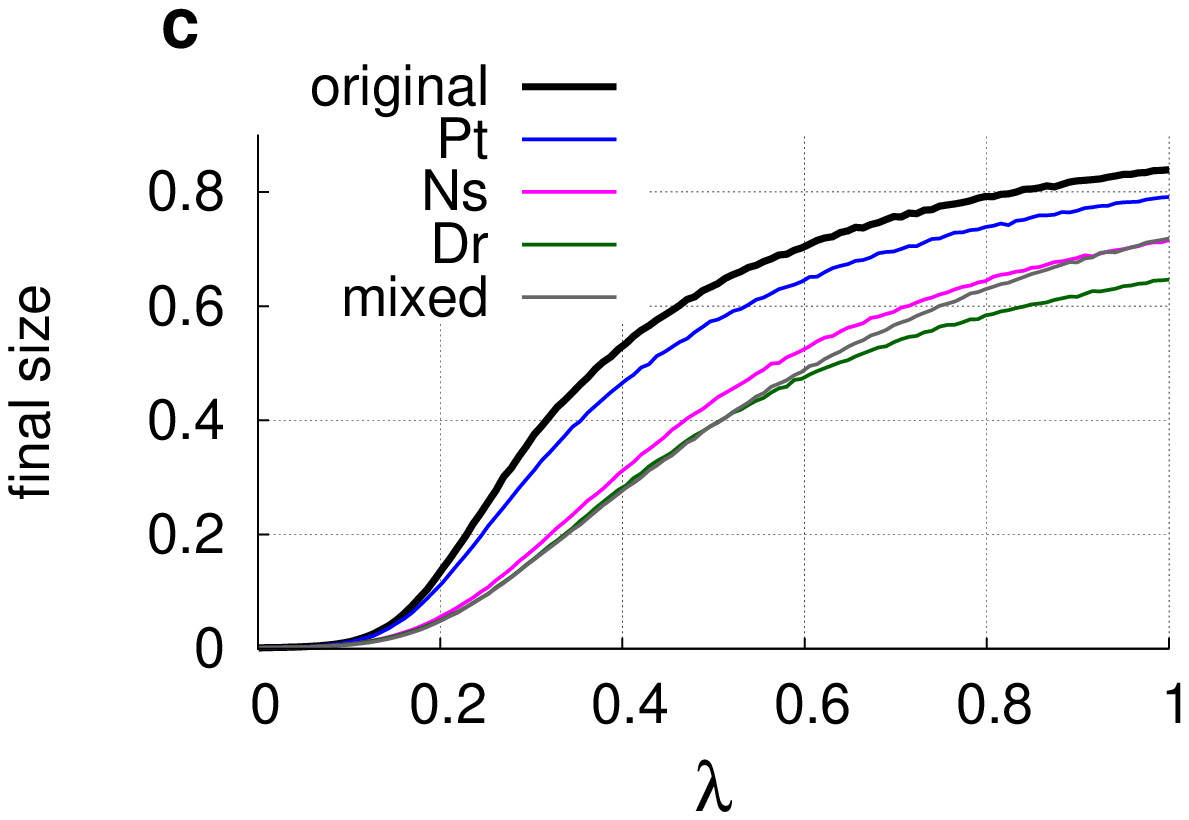}
\includegraphics[height=4cm,width=4cm]{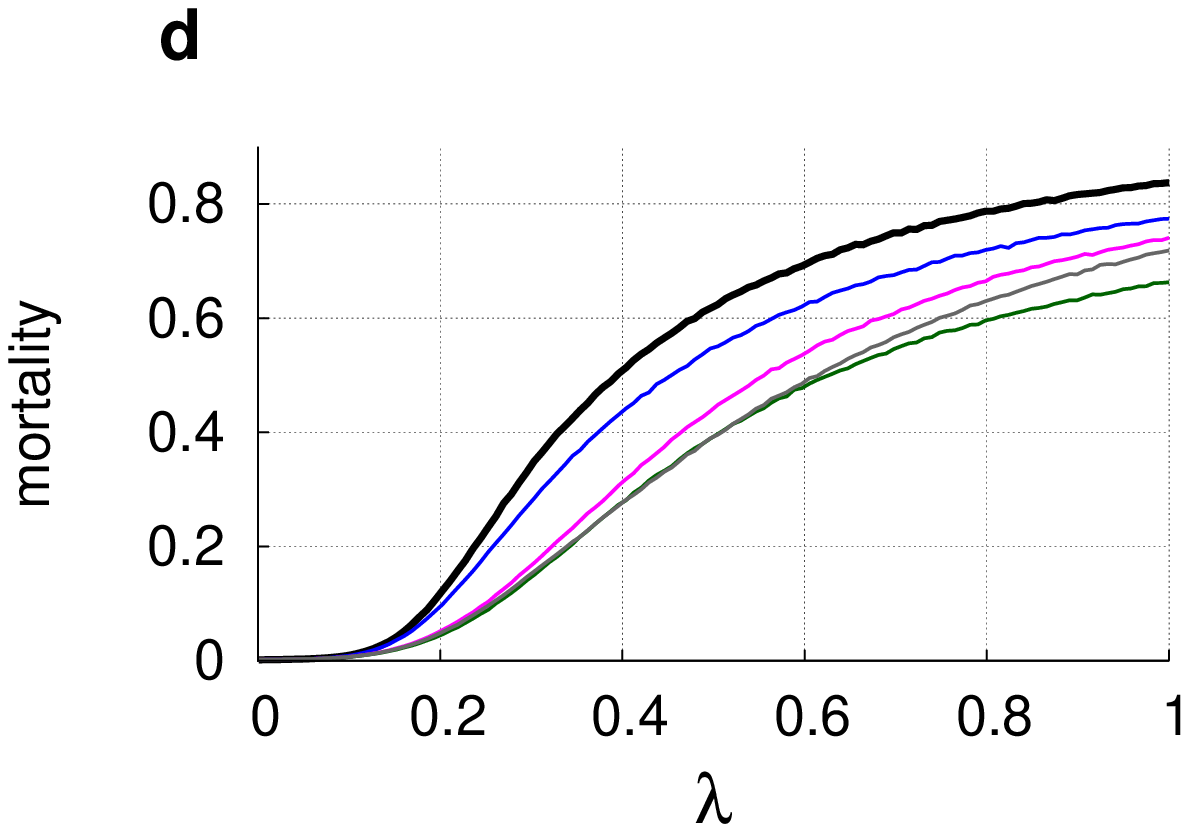}
\includegraphics[height=4cm,width=4cm]{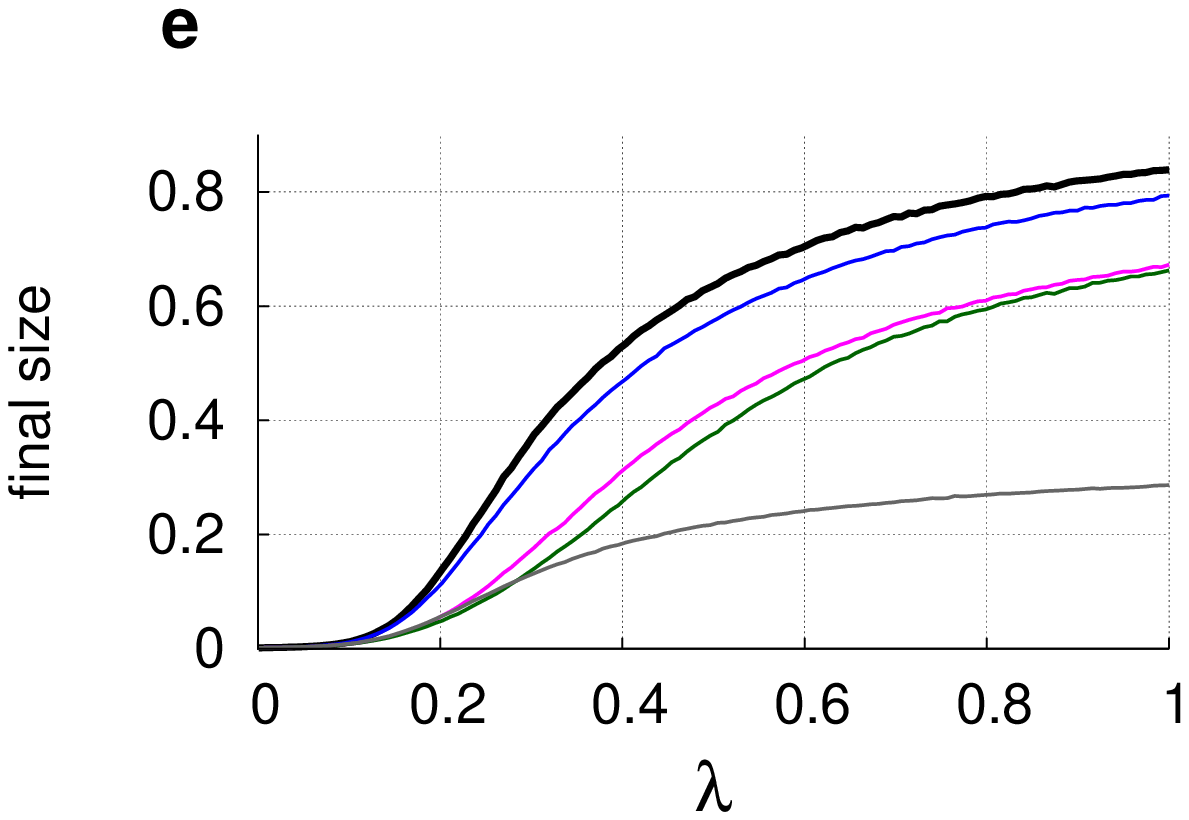}
\includegraphics[height=4cm,width=4cm]{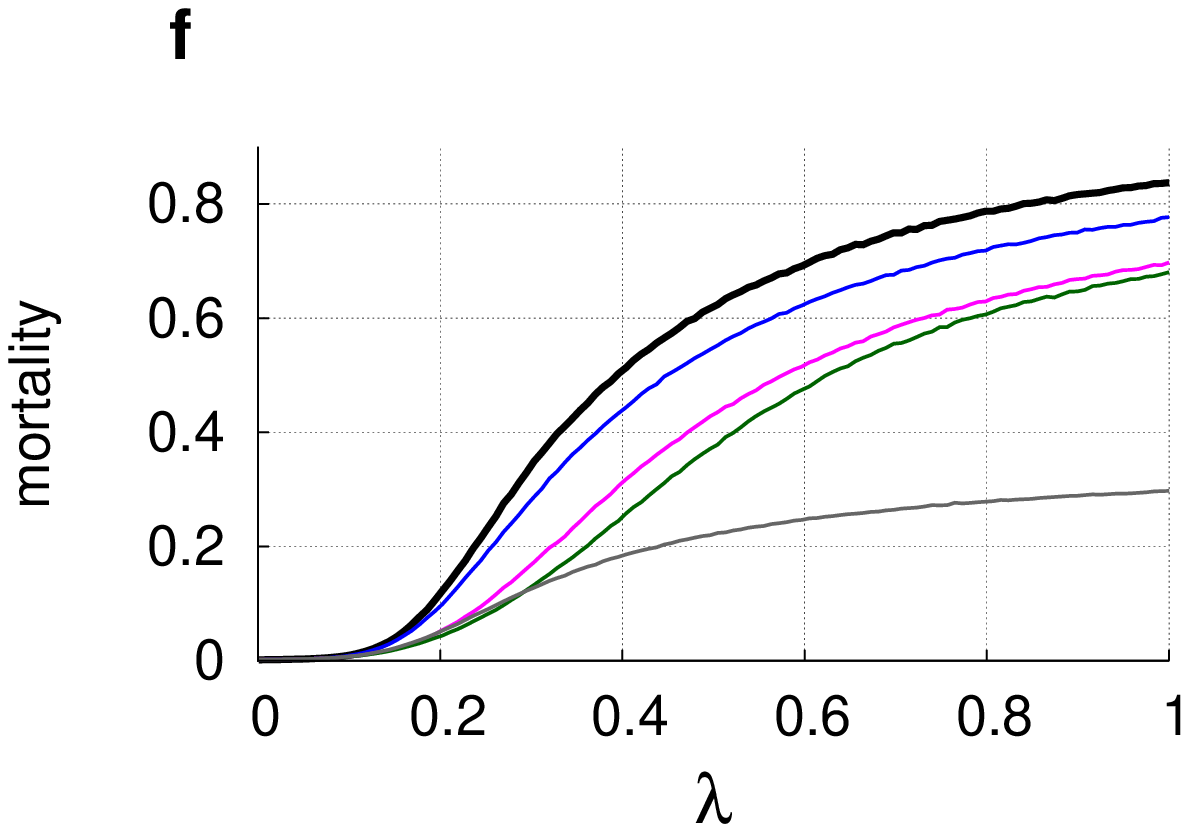}
\caption{Ueno and Masuda}
\label{fig:vac-day}
\end{center}
\end{figure}

Figure 6: Sensitivity analysis. The final size is compared for
different intervention protocols (see legends of
\FIG\ref{fig:r-inter-day} for details).  The infection rate is halved
for (a) susceptible nurses and doctors. (b) infected patients, and (c)
infected nurses and doctors.

\bigskip

Figure 7: Betweenness centrality plotted against
degree for the weekday
network for (a) patients, (b) nurses, and (c) doctors.

\bigskip

Figure 8: Effects of vaccination strategies for the weekday network.
(a) Final size and (b) mortality for different centrality-based and
random vaccination strategies. Different lines correspond to the
original observed network (`original'), degree-based vaccination
(`degree'), betweenness-based vaccination (`between'), vaccination
based on the recalculated betweenness centrality (`recal-betw'), and
random vaccination (`random'). (c) Final size and (d) mortality for
the vaccination strategy based on the betweenness centrality and the
class of individuals.  (e) Final size and (f) mortality for the
vaccination strategy based on the recalculated betweenness centrality
and the class of individuals. In (c)-(f), different lines correspond
to the original network (`original'), Pt-first protocol (`Pt'),
Ns-first protocol (`Ns'), Dr-first protocol (`Dr'), and class-mixed
protocol (`mixed').

\bigskip

Figure A1: Degree distributions of the weekend
network for different classes of individuals.

\bigskip

Figure A2: Disease spread in the weekend network.  (a) Final size for
the original network (solid lines) and the randomized network (dashed
lines).  (b) Distribution of the final size.  (c) Fraction of
mortality. (d) Distribution of the fraction of mortality.  For a fixed
$\lambda$, the upper masses of the distributions shown in (b) and (d)
actually consist of two peaks nearby. This is because just a single
junior doctor links the ward devoted to the psychiatry department with
39 (of 521) individuals to the rest of the network.

\bigskip

Figure A3: Effects of different intervention protocols for the weekend
network. (a) Final size and (b) fraction of mortality. Different
lines correspond to the original observed network (`original'), networks
after intervention 1 (`1'), intervention 1 with edge
duplication (`1 (dup)'), intervention 2 (`2'),
intervention 2 with edge duplication (`2 (dup)'), and
intervention 3 (`3'). See \TAB\ref{tab:weekend} 
for the number of edges for the networks before and after interventions.

\bigskip

Figure A4: Effects of vaccination strategies for the weekend network.
(a) Final size and (b) mortality for different centrality-based and
random vaccination strategies. Different lines correspond to the
original observed network (`original'), degree-based vaccination
(`degree'), betweenness-based vaccination (`between'), vaccination
based on the recalculated betweenness centrality (`recal-betw'), and
random vaccination (`random').  (c) Final size and (d) mortality for
the vaccination strategy based on the betweenness centrality and the
class of individuals.  (e) Final size and (f) mortality for the
vaccination strategy based on the recalculated betweenness centrality
and the class of individuals. In (c)-(f), different lines correspond to
the original network (`original'), Pt-first protocol
(`Pt'), Ns-first protocol (`Ns'),
Dr-first protocol (`Dr'), and class-mixed protocol (`mixed').

\bigskip

Table 1: The number of individuals and each type of edges for the
original weekday network and the networks after intervention.

\bigskip

Table A1: The number of individuals and each type of edges for the
original weekend network and the networks after intervention.

\clearpage

\clearpage

\clearpage

\clearpage

\clearpage

\clearpage

\clearpage

\clearpage

\clearpage

\renewcommand{\thefigure}{A\arabic{figure}}
\setcounter{figure}{0}

\begin{figure}
\begin{center}
\includegraphics[height=4cm,width=4cm]{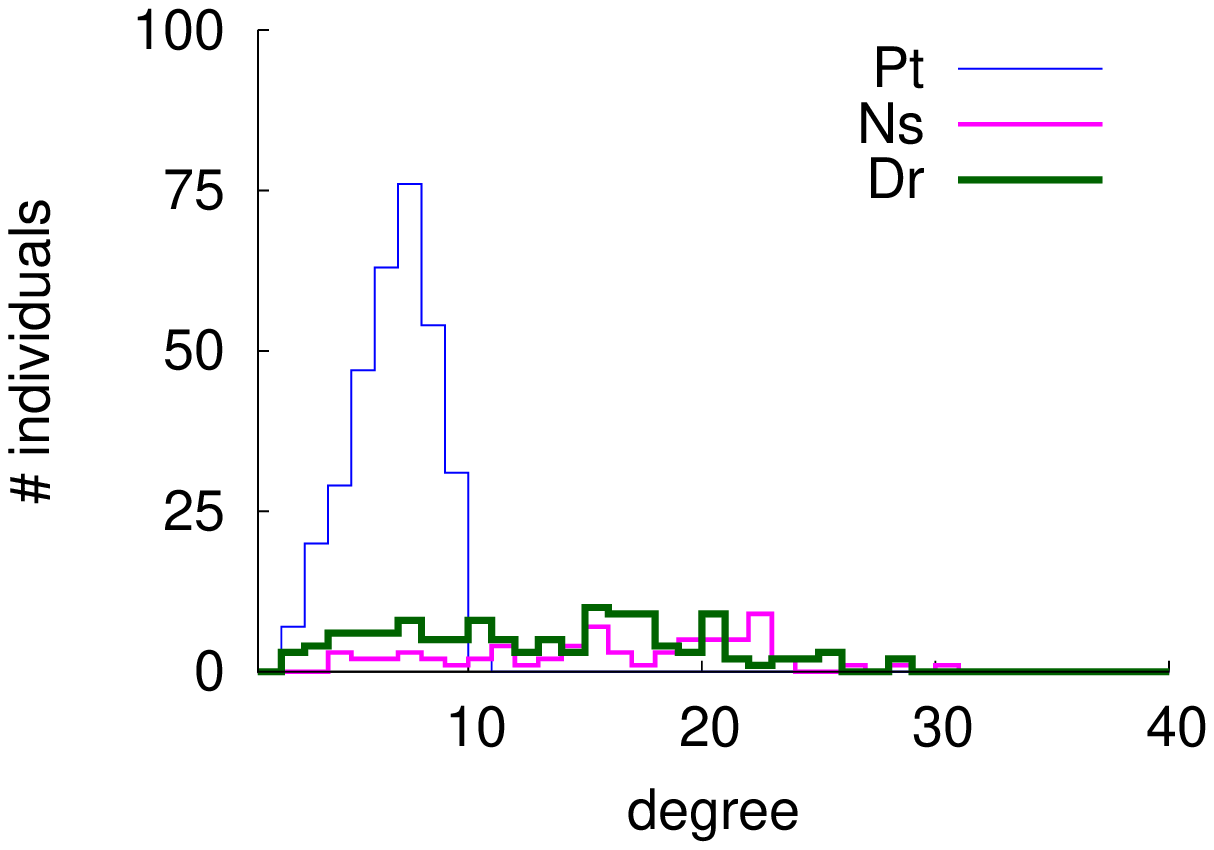}
\caption{Ueno and Masuda}
\label{fig:pk-end}
\end{center}
\end{figure}

\clearpage

\begin{figure}
\begin{center}
\includegraphics[height=4cm,width=4cm]{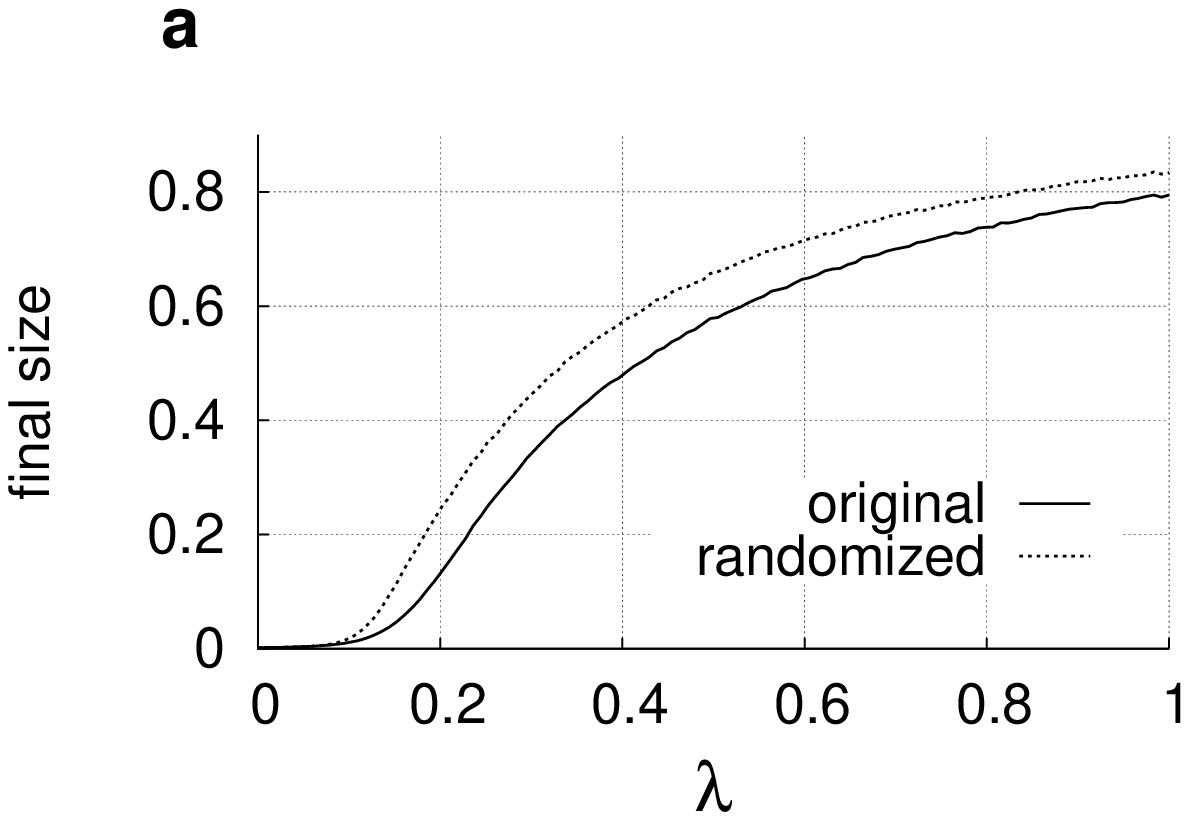}
\includegraphics[height=4cm,width=4cm]{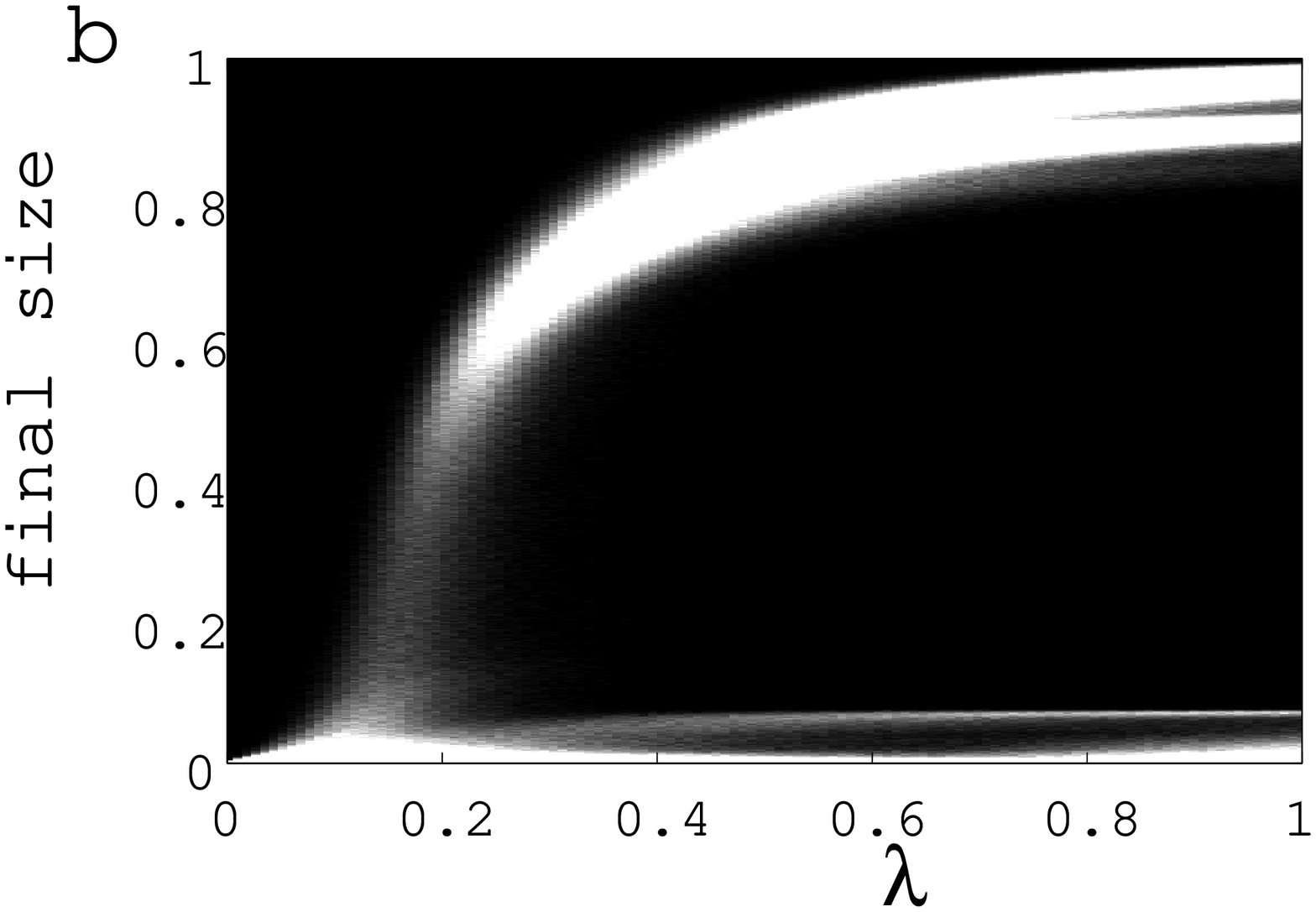}
\includegraphics[height=3cm,width=1.2cm]{final-morta-gauge.eps}
\includegraphics[height=4cm,width=4cm]{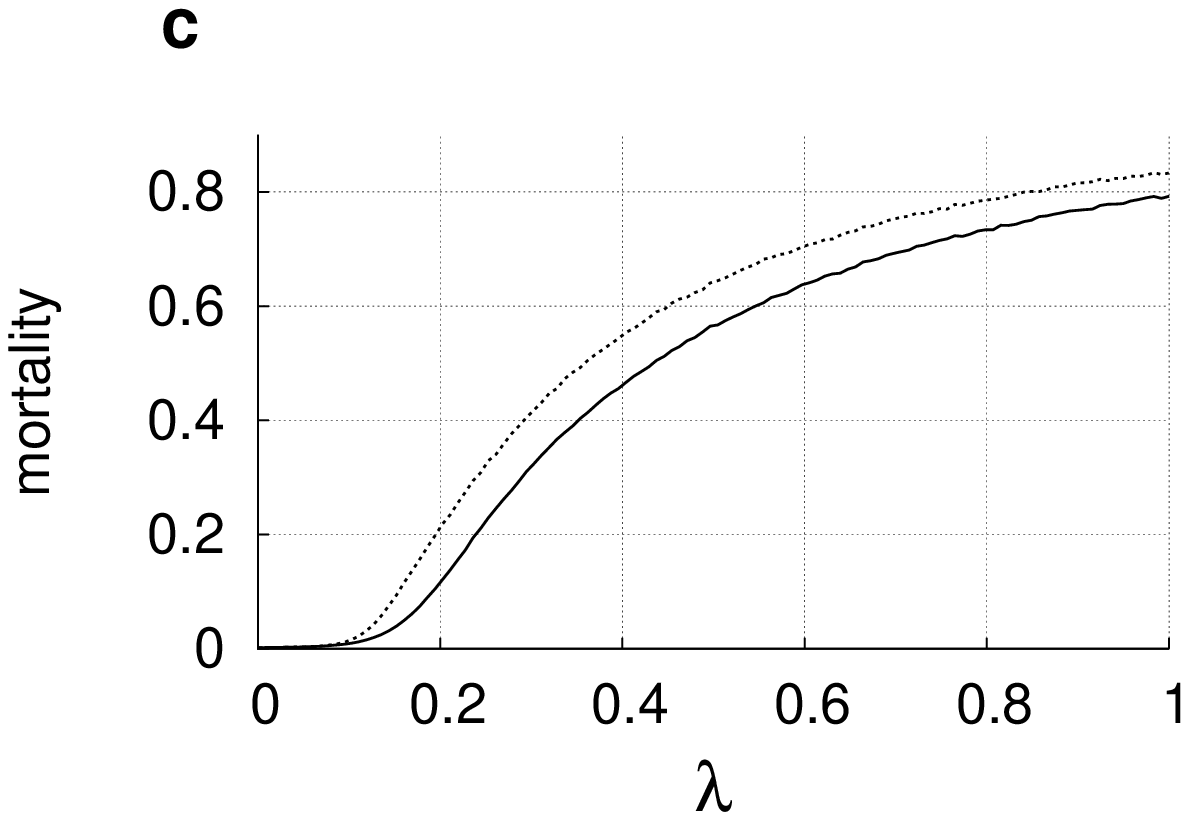}
\includegraphics[height=4cm,width=4cm]{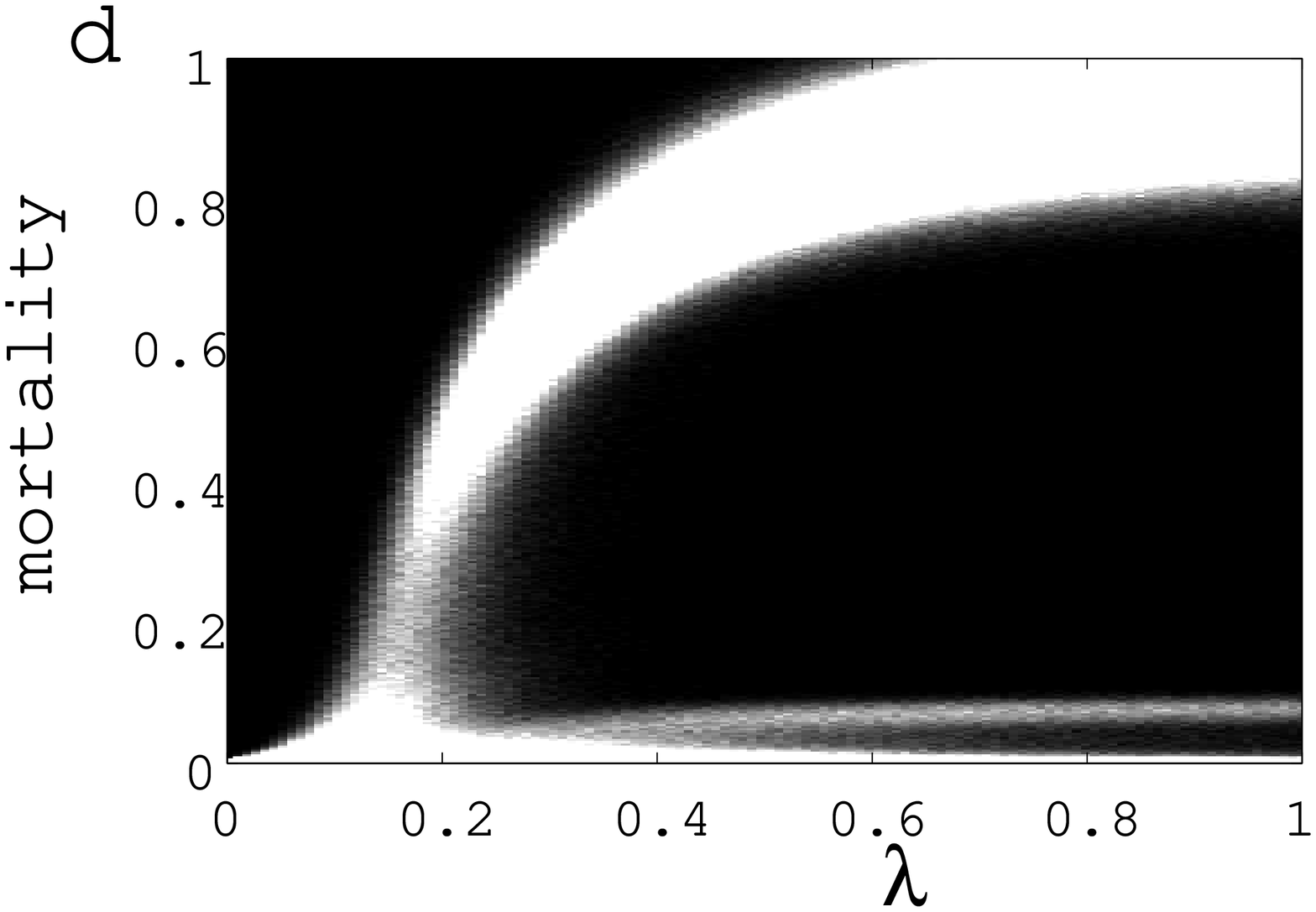}
\caption{Ueno and Masuda}
\label{fig:r-realrand-end}
\end{center}
\end{figure}

\clearpage

\begin{figure}
\begin{center}
\includegraphics[height=4cm,width=4cm]{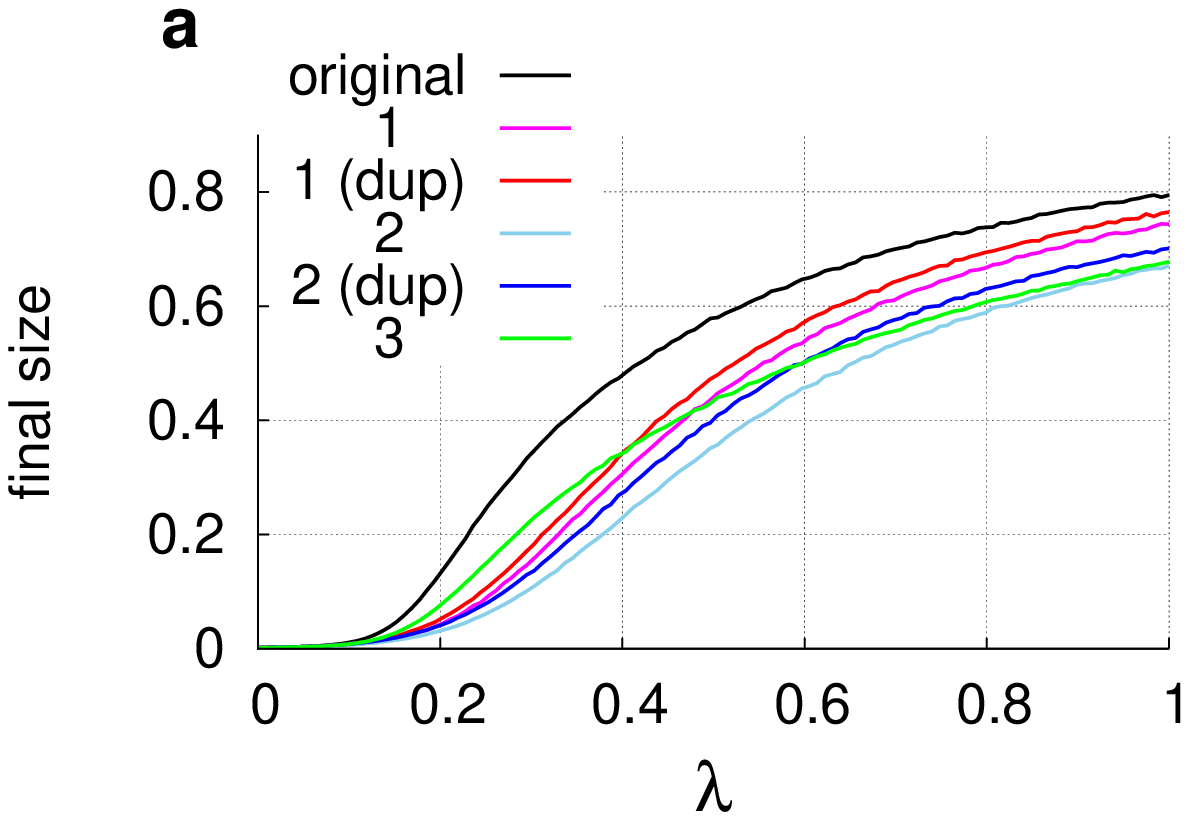}
\includegraphics[height=4cm,width=4cm]{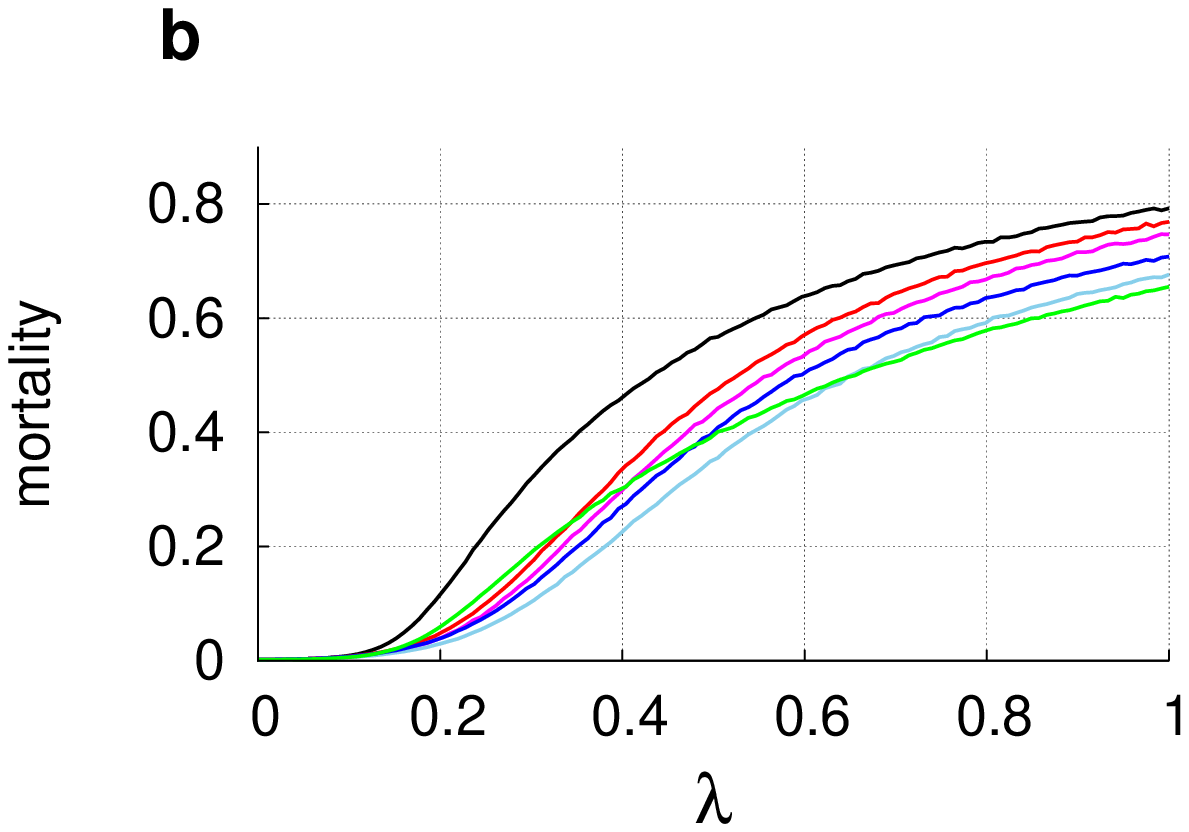}
\caption{Ueno and Masuda}
\label{fig:r-inter-end}
\end{center}
\end{figure}

\clearpage

\begin{figure}
\begin{center}
\includegraphics[height=4cm,width=4cm]{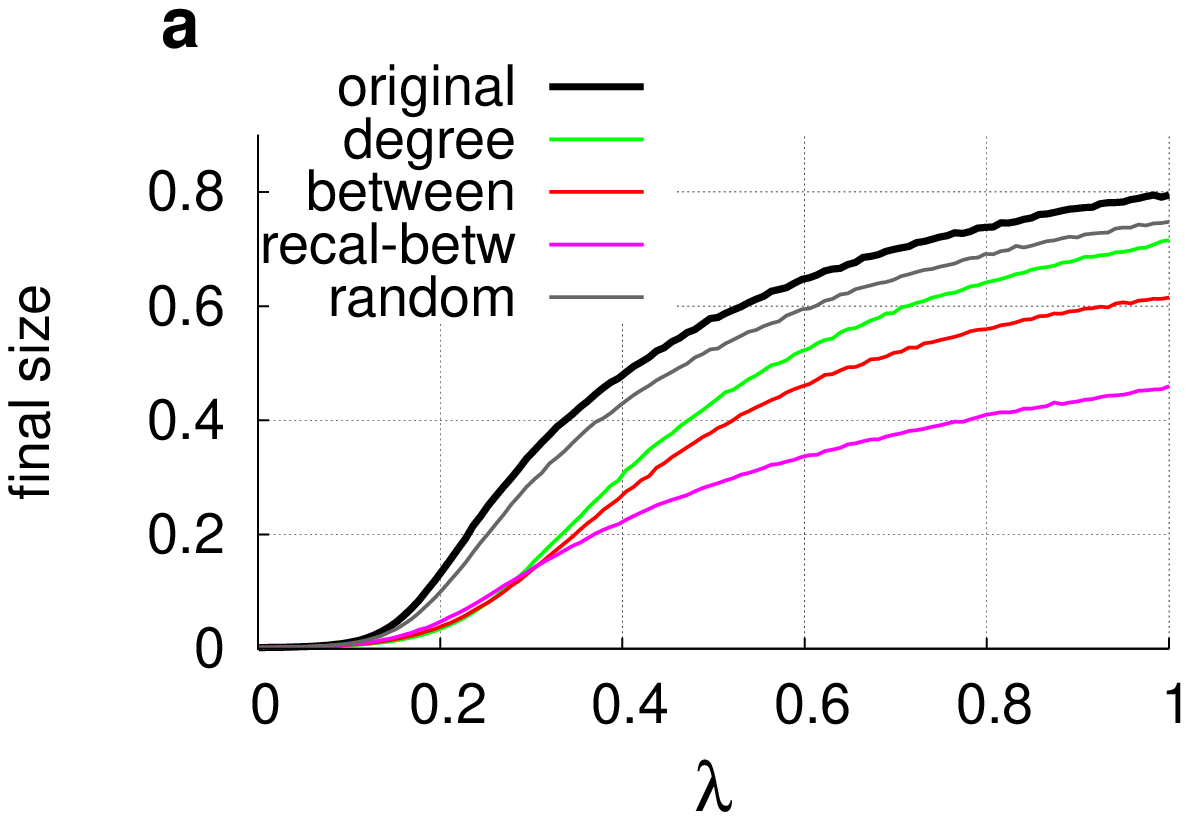}
\includegraphics[height=4cm,width=4cm]{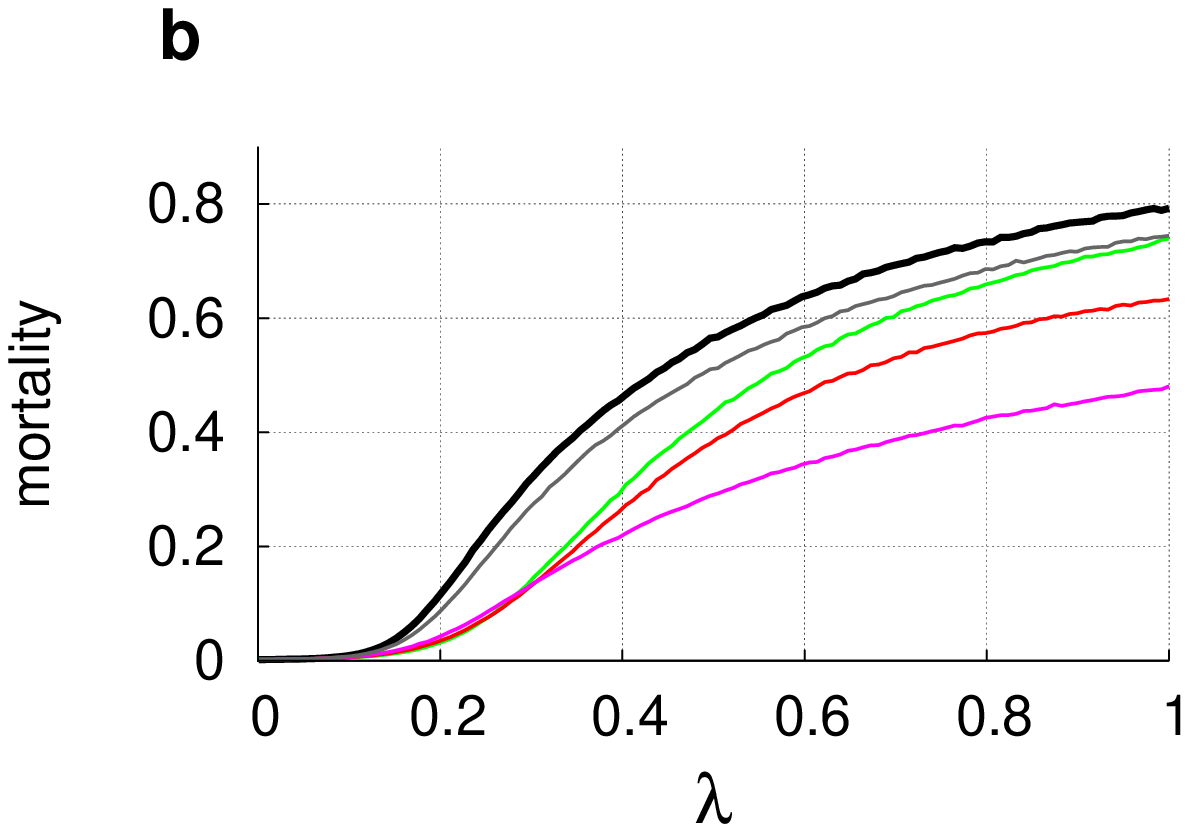}
\includegraphics[height=4cm,width=4cm]{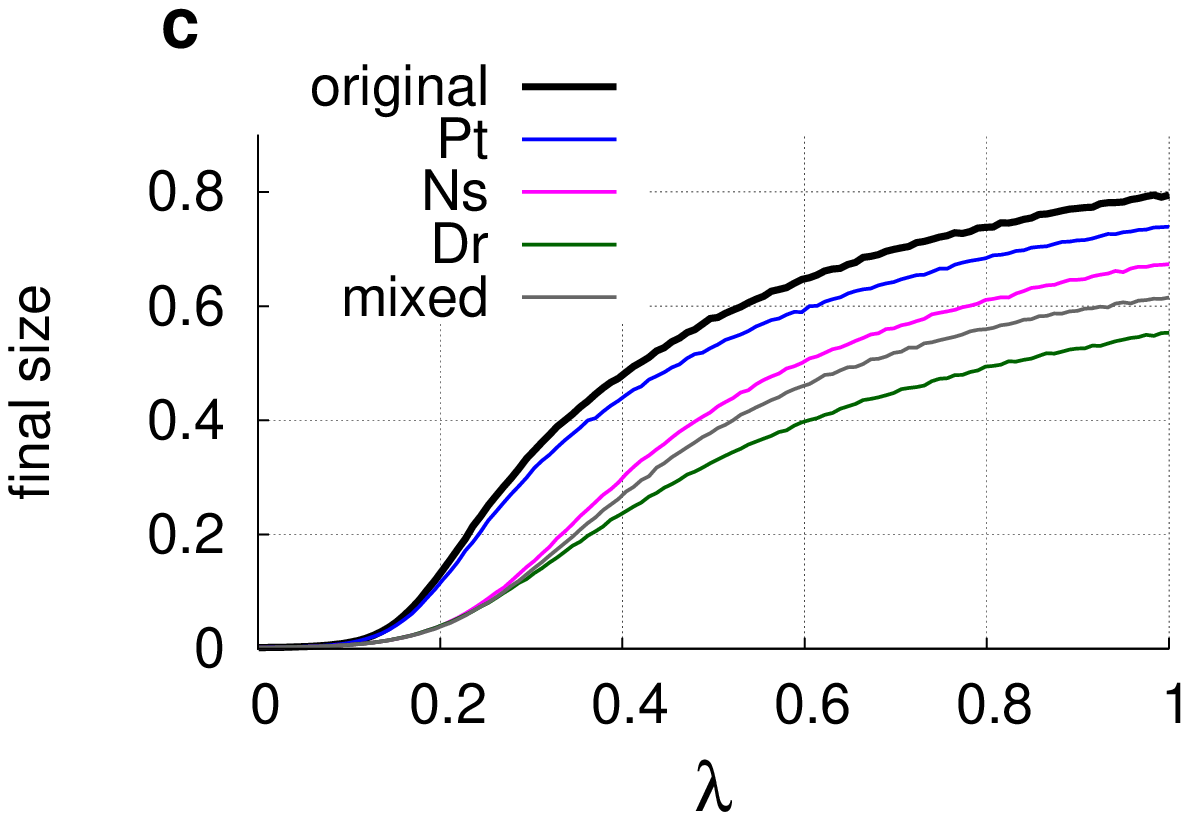}
\includegraphics[height=4cm,width=4cm]{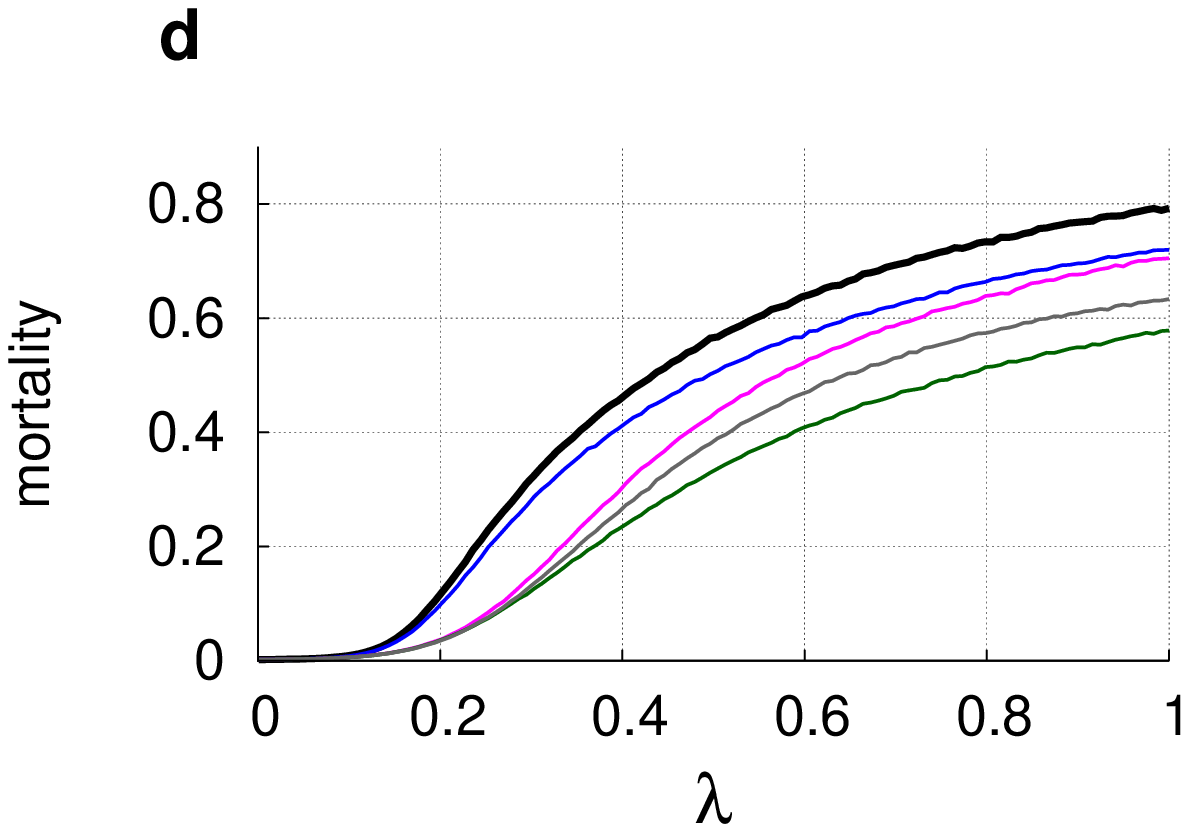}
\includegraphics[height=4cm,width=4cm]{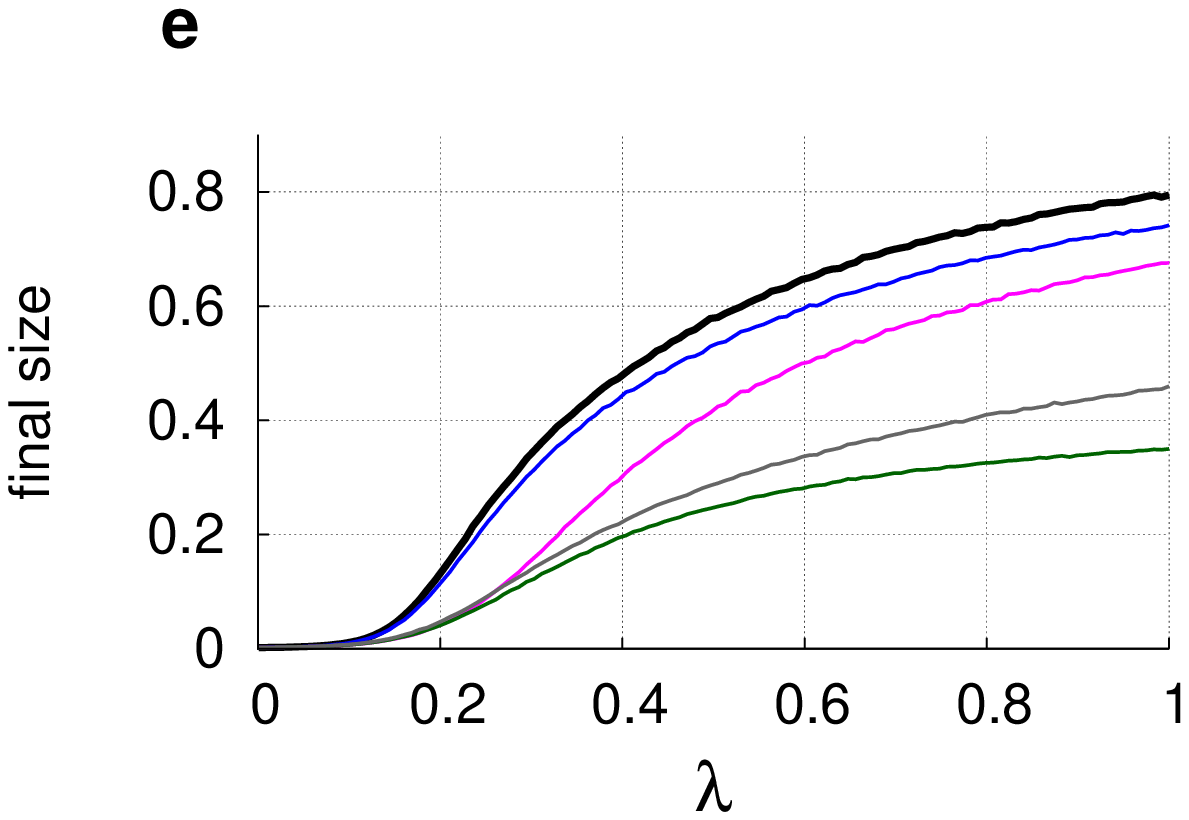}
\includegraphics[height=4cm,width=4cm]{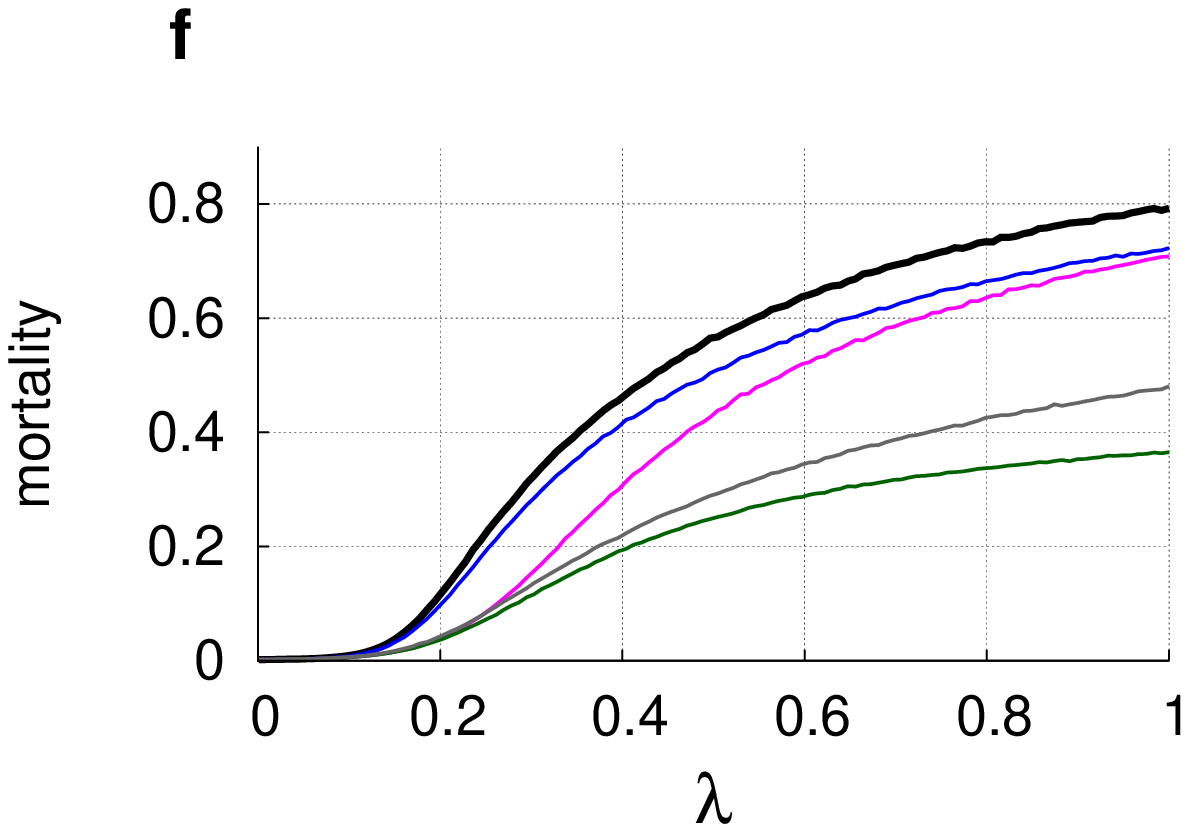}
\caption{Ueno and Masuda}
\label{fig:vac-end}
\end{center}
\end{figure}

\clearpage
\newpage

\begin{table}
\begin{center}
\caption{Ueno and Masuda}
\begin{tabular}{|p{2.4cm}|p{1.4cm}|p{1.4cm}|p{1.4cm}|p{1.4cm}|p{1.4cm}|p{1.4cm}|}
\hline
type of edge & original & 1 & 1 (dup) & 2 & 2 (dup) & 3\\ \hline
total & 3046 & 2823 & 3016 & 2717 & 3004 & 2382\\ \hline
Pt-Pt & 664 & 664 & 664 & 664 & 664 & 0\\ \hline
Pt-Ns & 388 & 388 & 388 & 388 & 388 & 388\\ \hline
Pt-Dr & 927 & 911 & 927 & 857 & 927 & 927\\ \hline
Ns-Ns & 245 & 245 & 245 & 245 & 245 & 245\\ \hline
Ns-Dr & 661 & 488 & 661 & 444 & 661 & 661\\ \hline
Dr-Dr & 161 & 131 & 131 & 119 & 119 & 161\\ \hline
\end{tabular}
\begin{tabular}{|c|c|}
\hline
type of individuals & number \\ \hline
total & 605\\ \hline
Pt & 388\\ \hline
Ns & 94\\ \hline
Dr & 123\\ \hline
\end{tabular}
\label{tab:weekday}
\end{center}
\end{table}

\clearpage
\newpage

\renewcommand{\thetable}{A\arabic{table}}
\setcounter{table}{0}

\begin{table}
\begin{center}
\caption{Ueno and Masuda}
\begin{tabular}{|p{2.4cm}|p{1.4cm}|p{1.4cm}|p{1.4cm}|p{1.4cm}|p{1.4cm}|p{1.4cm}|}
\hline
type of edge & original & 1 & 1 (dup) & 2 & 2 (dup) & 3\\ \hline
total & 2364 & 2168 & 2346 & 2094 & 2335 & 1872\\ \hline
Pt-Pt & 492 & 492 & 492 & 492 & 492 & 0\\ \hline
Pt-Ns & 329 & 329 & 329 & 329 & 329 & 329\\ \hline
Pt-Dr & 753 & 710 & 753 & 675 & 753 & 753\\ \hline
Ns-Ns & 124 & 124 & 124 & 124 & 124 & 124\\ \hline
Ns-Dr & 513 & 378 & 513 & 350 & 513 & 513\\ \hline
Dr-Dr & 153 & 135 & 135 & 124 & 124 & 153\\ \hline
\end{tabular}
\begin{tabular}{|c|c|}
\hline
type of individuals & number \\ \hline
total & 521\\ \hline
Pt & 329\\ \hline
Ns & 69\\ \hline
Dr & 123\\ \hline
\end{tabular}
\label{tab:weekend}
\end{center}
\end{table}

\end{document}